\documentclass[fleqn,usenatbib]{mnras}

\usepackage[T1]{fontenc}

\DeclareRobustCommand{\VAN}[3]{#2}
\let\VANthebibliography\thebibliography
\def\thebibliography{\DeclareRobustCommand{\VAN}[3]{##3}\VANthebibliography}

\usepackage{graphicx}	
\usepackage{amsmath}	
\usepackage{tabularx}
\usepackage{amssymb}	
\usepackage{newtxtext,newtxmath}
\usepackage{xcolor}
\usepackage{comment}
\usepackage{url}

\newcommand{\orcidicon}[1]{\href{https://orcid.org/#1}{\includegraphics[width=11pt]{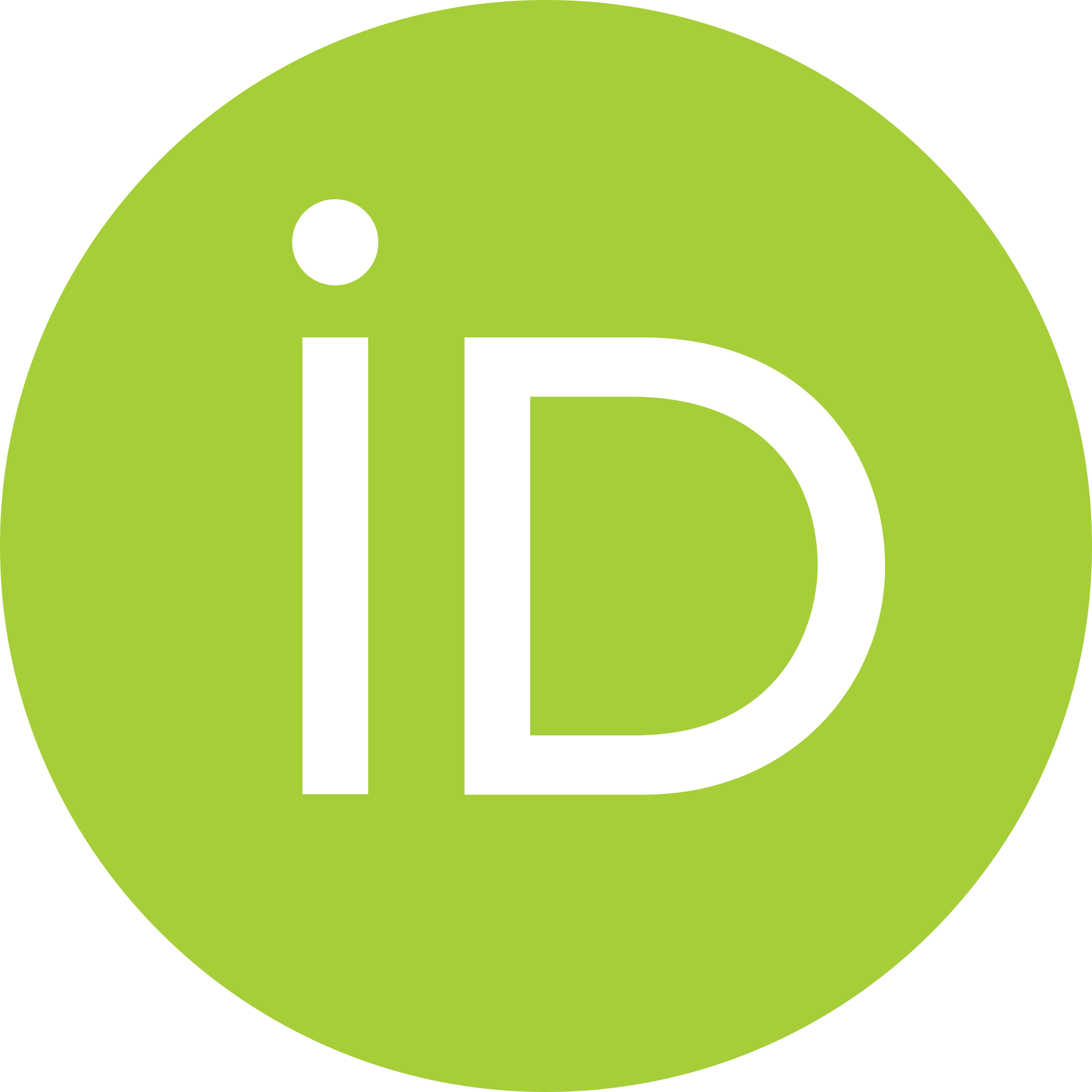}}}
\newcommand{\orcid}[1]{\href{https://orcid.org/#1}{\protect\orcidicon{#1}}}

\newcommand{\fehm}{\mathrm{[Fe/H]}}

\definecolor{gold}{rgb}{0.8, 0.5, 0.0}
\definecolor{purple}{rgb}{0.6, 0., 0.6}
\definecolor{darkgreen}{rgb}{0., 0.46, 0.26}
\definecolor{darkred}{rgb}{0.76, 0.23, 0.13}
\definecolor{darkblue}{rgb}{0.0, 0.0, 0.7}
\newcommand{\hanyuan}[1]{\textcolor{purple}{#1}}
\newcommand{\giu}[1]{\textcolor{darkgreen}{Giu: #1}}
\newcommand{\giuw}[1]{\textcolor{orange}{#1}}

\title[Young, metal-rich RR Lyrae]{Revealing the ages of metal-rich RR Lyrae via kinematic label transfer}

\author[Zhāng et al.]{HanYuan Zhāng\orcid{0009-0005-6898-0927},$^{1}$\thanks{hz420@cam.ac.uk}
Giuliano Iorio\orcid{0000-0003-0293-503X},\thanks{giuliano.iorio.astro@gmail.com}$^{2}$
Vasily Belokurov\orcid{0000-0002-0038-9584}$^{1}$,
N. Wyn Evans\orcid{0000-0002-5981-7360}$^{1}$, 
Alexey Bobrick\orcid{0000-0002-4674-0704}$^{3,4}$,
\newauthor
Valentina D'Orazi\orcid{0000-0002-2662-3762}$^{5,6,7}$
\\
$^{1}$ Institute of Astronomy, University of Cambridge, Madingley Road, Cambridge CB3 0HA, UK\\
$^{2}$ Departament de Física Quàntica i Astrofísica, Institut de Ciències del Cosmos, Universitat de Barcelona, Martí i Franquès 1, E-08028 Barcelona, Spain\\
$^{3}$ School of Physics and Astronomy, Monash University, Clayton, Victoria 3800, Australia\\
$^{4}$ ARC Centre of Excellence for Gravitational Wave Discovery -- OzGrav, Australia\\
$^{5}$ Department of Physics, University of Rome Tor Vergata, via della Ricerca Scientifica 1, 00133, Rome, Italy\\
$^{6}$ INAF Osservatorio Astronomico di Padova, vicolo dell'Osservatorio 5, 35122, Padova, Italy\\
$^{7}$ Fulbright Visiting Research Scholar, Department of Astronomy, The University of Texas at Austin, 2515 Speedway, Austin, TX 78712, USA
}

\date{Accepted XXX. Received YYY; in original form ZZZ}

\pubyear{2025}

\begin{document}
\label{firstpage}
\pagerange{\pageref{firstpage}--\pageref{lastpage}}
\maketitle

\begin{abstract}
RR Lyrae stars have long been considered reliable tracers of old, metal-poor populations, primarily due to their prevalence in globular clusters and the Galactic halo. However, the discovery of a metal-rich subpopulation in the Galactic disc, kinematically colder and more rotationally supported, challenges this classical view. Understanding the age of these metal-rich RR Lyrae stars is crucial for constraining their formation pathways and assessing what Galactic populations they are tracing. In this work, we leverage the unprecedented astrometric precision of Gaia DR3 to infer the age distribution of metal-rich RR Lyrae stars through a kinematic comparison with O-rich Mira variables. Mira variables, with their well-established period-age relation, serve as a natural clock, allowing us to transfer age information to RR Lyrae stars via their phase-space properties. 
By applying this approach across different metallicity bins, we find that the most metal-rich RR Lyrae stars ($[\rm Fe/H] > -0.5$) exhibit kinematics consistent with long-period ($\rm{period}\approx 150\,\rm{days}$), young Mira variable population; its age corresponds to $\sim 6-7$ Gyr (adopting the period-age relation in \citealt{ZS23}) that is significantly younger than typically assumed for RR Lyrae stars. In contrast, those with $-1 < [\rm Fe/H] < -0.5$ show properties more aligned with older ($\approx 9-11$ Gyr) populations. Interestingly we also find evidence of a possible double age populations for the most metal-rich RR Lyrae, one younger with ages between 4 and 6 Gyr, and another one older ranging from 8 to 9 Gyr. These results provide strong evidence that metal-rich RR Lyrae stars in the Galactic field do not exclusively trace ancient populations. This finding challenges the current model of RR Lyrae formation and supports alternative formation scenarios, such as binary evolution.
\end{abstract}




\section{Introduction}

RR Lyrae stars are among the most studied variable stars, due to their unique combination of observational accessibility and astrophysical significance \citep{Smith_2004,Catelan_2009,Catelan_2015}. They pulsate in the fundamental (RRab) mode, first-overtone mode (RRc) or a mix of the two (RRd) with periods ranging between 0.2 and 1.0 days.  RR Lyrae are helium-burning low-mass stars, hot enough (5500-7500 K) to cross the classical instability strip while evolving on the Horizontal Branch \citep{Smolec_2008,Marconi_2015,DeSomma_2024}. Models of stellar evolution show that metal-poor ([Fe/H]<-1) stars with initial mass between 0.8-1.0 Msun can efficiently become RR Lyrae after 10 Gyr \citep{Bono_1994, Bono_1995}.  RR Lyrae are indeed abundant amongst old and metal-poor populations typical of the stellar halo and the globular clusters (so-called Population II). Their narrow colour range, relatively high brightness  and the existence of a tight relation between the light curve properties, their metallicity and the luminosity (PLZ relations, see e.g.\ 
\citealt{Dekany_2022,Li_2023,Mullen_2023,Prudil_2024,Muraveva_2025}) have made RR Lyrae an extraordinary tool to anchor distances in the Milky Way \citep[e.g.][]{Garofalo_2022,Muraveva_2024}, globular clusters and the Local Group \citep[e.g.][]{Clementini_2001, Bono_2019, Savino_2022, Garofalo_2024}, and to study the properties, structure and sub-structure of the Galactic innermost part \citep[e.g.][]{Du_2020,Kunder_2020,Savino_2020,Han_2024,  Prudil_2025}, Galactic stellar halo \citep[e.g.][]{Belokurov_2018,Iorio_Belokurov_2019,Simion_2019,Li2022,Ablimit_2022,Cabrera_2024}, stellar streams \citep[e.g.][]{Duffau_2006,Mateu_Stream_2018,Price_2019,Coppi_2024}, Galactic disk \citep[e.g.][]{Mateu_2018,Cabrera_2024_warp}, and nearby Galaxies such as the Magellanic Clouds \citep[e.g.][]{Belokurov_2017,Muraveva_LMC_2018,Cusano_2021,Cuevas_2024}, M31 and M33 \citep[e.g.][]{Sarajedini_2006,Pritzl_2011,Tanakul_2018}. 
RRLs have also been employed as distance indicators in the calibration of the cosmological distance ladder for nearby galaxies, complementing measurements based on brighter Cepheid variables \citep[see e.g.][]{Beaton_2016,Riess_2016}.

Although RR Lyrae stars are traditionally associated with metal-poor populations, the presence of a metal-rich subgroup, extending up to solar and super-solar metallicities, in the Solar neighbourhood has been known since the pioneering work of \cite{Preston_1959} which was later confirmed by \cite{Tam_1976} and \cite{Zinn_1985}.
The first comprehensive chemo-dynamical studies of RR Lyrae stars in the solar neighbourhood revealed a clear kinematic transition at $\fehm=-1$, where their orbits and spatial distributions shift from a halo-like to a disk-like configuration \citep{Layden_1994,Layden_1995a, Layden_1995b}. For $\fehm\geq-0.5$, RR Lyrae stars appear to settle into a colder, more rotationally supported, and thinner disk.
As astrometric and spectroscopic observations of RR Lyrae stars in the Solar neighbourhood have improved, subsequent studies have confirmed and refined our understanding of the metal-rich RR Lyrae population in both the Milky Way and the Magellanic Clouds \citep{Soszynski_2016,Prudil_2020,Crestani_2021,Clementini_2023,DOrazi_2024,Prudil_2025}. Using the unprecedented all-sky catalogues of RR Lyrae stars provided by the Gaia satellite \citep{GaiaDR2,Clementini_2019}, \cite{IB21} extended the chemo-kinematic analysis well beyond the solar vicinity. Their results confirmed that RR Lyrae stars with velocity dispersions characteristic of intermediate-young populations (2–5 Gyr; \citealt{Sharma_2021}) exist throughout the entire Milky Way thin disk (up to 15-20 kpc). Further leveraging Gaia data, \cite{Cabrera_2024_warp} showed that the metal-rich RR Lyrae population traces the Galactic warp similarly to 3-4 Gyr-old thin disk tracers. Additionally, \cite{Sarbadhicary_2021} and \cite{Cuevas_2024} found evidence linking RR Lyrae stars to intermediate-young (1–8 Gyr) stellar populations in the Magellanic Clouds.

These studies challenge the traditional view of RR Lyrae stars as exclusive tracers of old stellar populations. In particular, they suggest that in the Milky Way, most metal-rich RR Lyrae stars share the spatial and kinematic properties of Population I thin-disk stars with ages ranging from 2 to 10 Gyr. Chemically, these metal-rich RR Lyrae stars are $\alpha$-poor \citep{Marsakov_2018,Crestani_2021}, consistent with the typical thin disk population. However, they exhibit low abundances of specific elements, such as yttrium, scandium, and barium \citep{Gozha_2021,DOrazi_2024,Gozha_2024}, a feature more commonly associated with older stellar populations \citep[see e.g.][]{Sheminova_2024}.

As first noted by \cite{Tam_1976}, the existence of such a population in the thin disc poses a fundamental conundrum. At high metallicities ($\fehm>-1$), increased envelope opacity causes low-mass core-helium-burning stars to be cooler than their metal-poor counterparts of the same mass. For $\fehm>-0.5$, only stars with masses around 0.5 M$_\odot$ in the core-helium-burning stage have an envelope thin enough to enter the instability strip. To form an RR Lyrae star under these conditions, the initial mass must be low ($\lesssim0.7$~M$_\odot$) and/or the stars must lose a significant amount of mass ($\gtrsim 0.5$~M$_\odot$) during their red giant branch (RGB) ascent \citep{Tam_1976,Bono_97a}.
In the first scenario, stellar evolution models predict that such stars would be among the oldest in the Milky Way, possibly older than the Hubble time \citep[see e.g.][]{Lee_1992,Savino_2020}. In the second scenario, the initial mass could be larger (up to 2 M$_\odot$), allowing classical thin disc Population I stars to evolve into RR Lyrae stars \citep{Strugnell_1986,Bono_97b}. However, the required amount of mass loss is inconsistent with the wind mass-loss rates derived from studies of RGB and horizontal branch stars in globular clusters, dwarf spheroidals, and field stars \citep[$\lesssim 0.3~\mathrm{M}_\odot$][]{Gratton_2010,Origlia_2014,Savino_2019,Tailo_2020,Tailo_2021,Miglio_2021,Brogaard_2024,Tailo_2025}.
Thus, alternative formation channels are required. Proposed alternatives include helium enrichment \citep[see e.g.][]{Lee_2016,Savino_2020,Gozha_2024}, unusually high mass loss due to binary interactions \citep[see e.g.][]{Pietrzynski_2012}, or dynamical interactions in dense environments \citep[see e.g.][]{FusiPecci_1993,Pasquato_2013}. In particular, \cite{Karczmarek_2017} and \cite{Bobrick2024} found that mass transfer in binary systems can produce metal-rich RR Lyrae stars with ages consistent with those of the thin disc.
However, to date, only two RR Lyrae stars have been confirmed in binary systems \citep{Pietrzynski_2012,Liska_2016}, although a list of potential candidates spans the range of periods and masses predicted by binary formation models \citep{Hajdu_2021}.
The observed peculiar under-abundances of certain elements (yttrium, scandium, and barium) could support the binary origin, similar to post-AGB or post-RGB stars \citep[see e.g.][]{Kamath_2015,Mohorian_2024,Mohorian_2025}. Alternatively, these stars may not be a genuinely intermediate-young thin-disc population but rather an accreted population \citep[see e.g.][]{Feuillet_2022,DOrazi_2024} or stars that migrated from the inner regions of our Galaxy \citep[see e.g.][]{Frankel2020,Lu_2024,Marques_2025,Zhang_2025}.

Constraining the ages of metal-rich RR Lyrae stars is essential for distinguishing between different formation scenarios. This has significant implications not only for the study of RR Lyrae stars but also for our understanding of stellar and binary evolution, as well as the formation and evolution of our Galaxy.
In this work, we aim to constrain the age of this intriguing RR Lyrae population by comparing its phase-space distribution with that of another population of variable stars—the Mira variables—whose ages can be inferred from their pulsation periods.

Mira variables are long-period pulsating stars in the asymptotic giant branch (AGB) phase of stellar evolution, characterised by large amplitudes in brightness and periods ranging from 80 to 1000 days \citep{Matsunaga2009}. Even though they are typically referred to as an intermediate-age population, they have a wide age range: from 1 to about 12 Gyr. A crucial characteristic of Mira variables is the correlation between their pulsation period and ages, which has been demonstrated in many studies \citep{Feast2000, Clement2001, FeastWhitelock2014, Grady2019, TrabucchiMowlavi2022, ZS23}. Due to this special feature, Mira variables are commonly used in Galactic modelling \citep[e.g.][]{Catchpole2016, Semczuk2022, Sanders2024, Zhang_2024}. We adopt the period-age relation in \citet{ZS23}, which is kinematically calibrated using astrometric measurements from \textit{Gaia} DR3. By comparing the phase-space properties of Mira variables and RR Lyrae, we can leverage this period-age relationship to place constraints on the ages of RR Lyrae stars as a function of metallicity. 

The paper is organised as follows. In Section \ref{sec:data} we present the RR Lyrae and Mira dataset used in this work. In Section \ref{sec::methodology}, we present the method used to compare the two datasets and retrieve the RR-Lyrae ages, and we also present the test performed on Cosmological simulations. Section \ref{sec:results} reports the results of our analysis, while  in  Section \ref{sec:discussion} we discuss their implications and the caveat/limitations of our analysis. Finally, in Section \ref{sec:conclusions}, we summarise the main take-aways of this paper.

\begin{figure*}
    \centering
    \includegraphics[width=\textwidth]{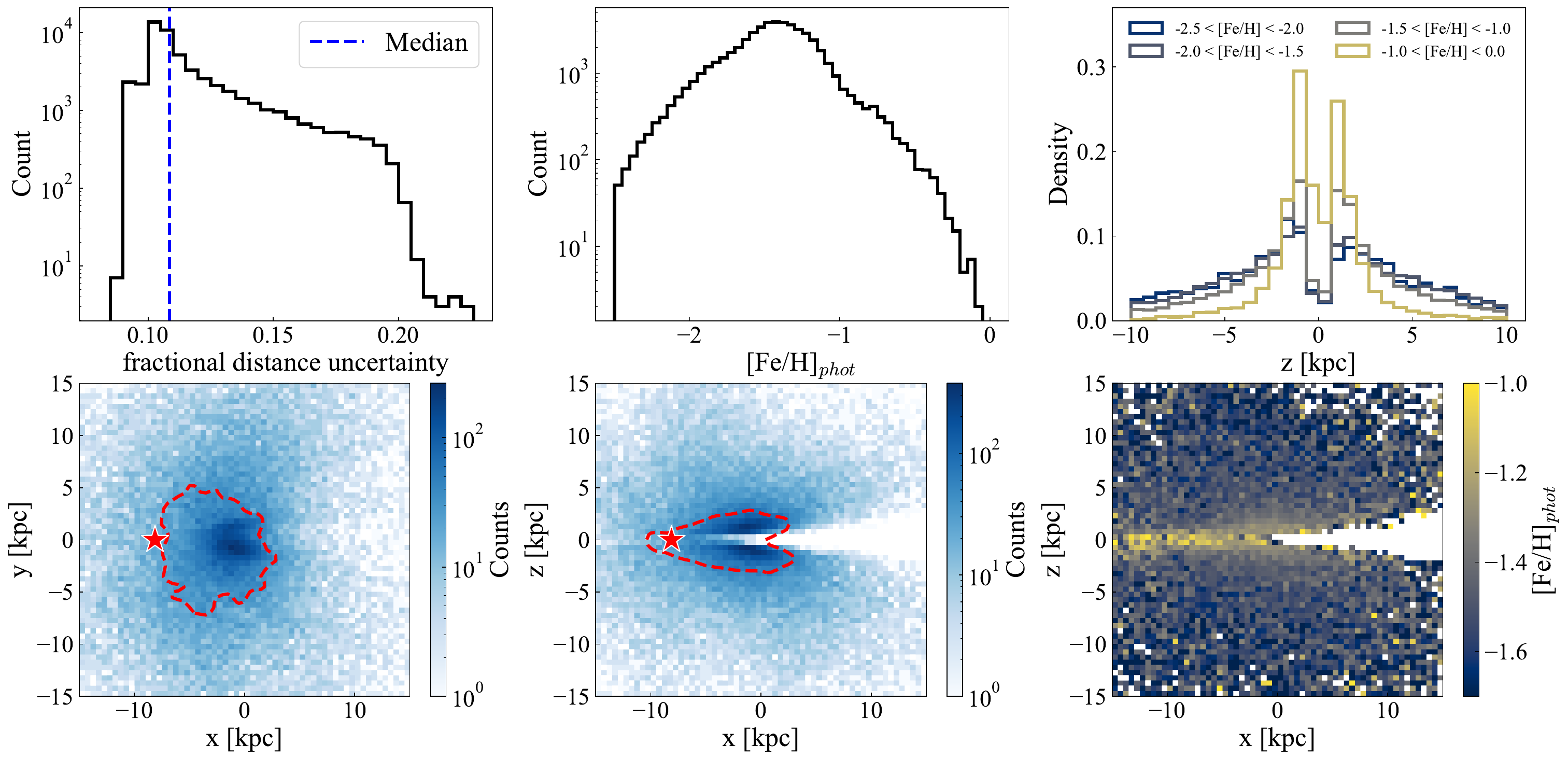}
    \caption{Properties of the RRL sample. {\it Top left:} distribution of the fractional distance uncertainty. {\it Top middle:} metallicity distribution. {\it Top right:} $z$-distribution of RRL in various metallicity bins. {\it Bottom left:} the spatial face-on ($x$-$y$) distribution of RRL, where the red contours enclosed $75\%$ of the RRL with ${\rm [Fe/H]}>-1$. The red star with a white edge labels the location of the Sun in this coordinate. {\it Bottom middle:} the spatial edge-on ($x$-$z$) distribution of the RRL candidates, where the red contours enclosed $75\%$ of the RRL with ${\rm [Fe/H]}>-1$. {\it Bottom right:} the mean metallicity of RRLs in the $x$-$z$ plane.}
    \label{fig:RR-Lyrae_info}
\end{figure*}

\section{Data} \label{sec:data}
\label{sec::data}
\subsection{RR-Lyrae sample}
\label{sec::data:RR}

\begin{figure}
    \centering
    \includegraphics[width=\columnwidth]{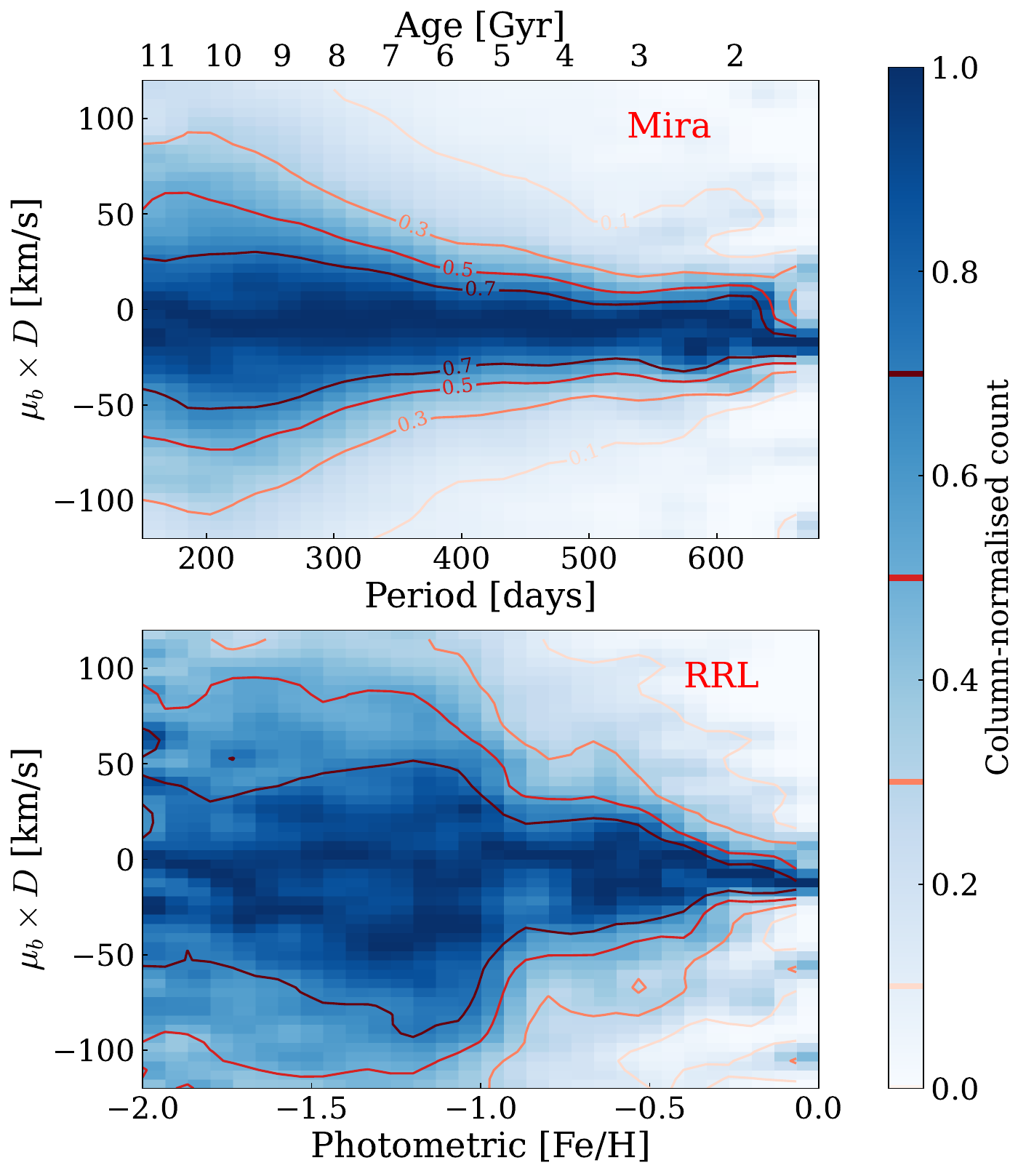}
    \caption{{\it Top panel:} the column-normalised Galactic latitude velocity$-{\rm age}$ distribution of the Mira variable sample. {\it Bottom panel:} the same column-normalised distribution of the RRLs that reside within $2$~kpc above or below the Galactic plane. }
    \label{fig:columnnormalised_vb}
\end{figure}

We adopt the cleaned RR Lyrae stars (RRLs) sample from \citet{IB21} with updated {\it Gaia} DR3 light curves and astrometry (see \citealt{Bobrick2024}). \citet{IB21} followed a similar procedure to \citet{Iorio_Belokurov_2019} to clean the RRL catalogue of \textit{Gaia} DR2 \citep{Clementini_2019,GaiaDR2}. RRLs associated with compact structures in our Galaxy (e.g., globular clusters and dwarf galaxies) are removed. RRLs with bad astrometry and photometry are also removed using cuts on extinction, BP/RP colour excess, and renormalised unit weight error (RUWE). The metallicity of RRLs is correlated to the characteristic pulsation period and the phase difference of the third and the first harmonics, $\Phi_{31}$. We adopt the photometric metallicity from \citet{IB21} and remove RRLs if their $\Phi_{31}$ is unavailable (see more details in \citealt{IB21}). There are 53 165 RRL candidates with photometric metallicity remaining. The luminosity distance is assigned using the $M_G-\mathrm{[Fe/H]}$ relation in \citet{Muraveva_2018}, where $M_G$ is the absolute magnitude in the $G$-band. The distributions of the fractional distance uncertainty and the photometric metallicity of the RRLs are shown in the top left and middle panels of Fig.~\ref{fig:RR-Lyrae_info}. The median fractional distance uncertainty of the RRL sample is 11$\%$. The top right panel shows the $z$-distribution of RRLs in different metallicity bins; the metal-rich RRLs ([Fe/H]$_\mathrm{phot}$ > -1) stay closer to the plane compared to the metal-poor RRLs (see also \citealt{Bobrick2024}). Correspondingly, the mean metallicity of the RRLs is higher when they are closer to the Galactic mid-plane, as illustrated in the lower right panel. The bottom left and middle panels show the spatial distribution of the RRL sample in the $x$-$y$ and $x$-$z$ planes. In this work, we focus the analysis on RRL with metallicity of [Fe/H]$_\mathrm{phot}$ > -1, which we refer to as the metal-rich RRL candidates in later discussions. Among the selected 53 165 RRL candidates, 3224 have [Fe/H]$_\mathrm{phot}$ > -1. The kinematic evolution of RRL as a function of metallicity is shown in the lower panel of Fig.~\ref{fig:columnnormalised_vb} in terms of the Galactic latitudinal velocity $\mu_b\times D$, where $\mu_b$ is the proper motion in the Galactic latitude direction, and $D$ the heliocentric distance. The velocity dispersion drops significantly for RRL with $[\rm Fe/H]>-1$. 

\begin{figure*}
    \centering
    \includegraphics[width=\textwidth]{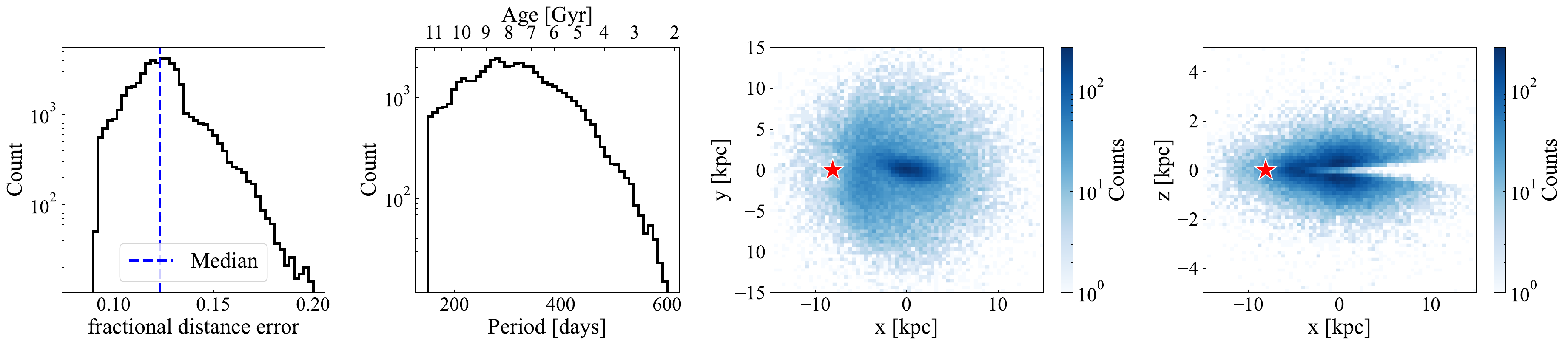}
    \caption{Properties of the O-rich Mira sample. {\it Leftmost:} fractional distance uncertainty distribution. {\it Middle left:} Period (age) distribution of the O-rich Mira variable sample. The age is computed from the characteristic period using the period-age relation in \citet{ZS23}. {\it Middle right:} the spatial face-on ($x$-$y$) distribution of the O-rich Mira candidates. The red star with a white edge labels the location of the Sun in this coordinate. {\it Rightmost:} the spatial edge-on ($x$-$z$) distribution of the O-rich Mira candidates.}
    \label{fig:Mira_info}
\end{figure*}


\subsection{O-rich Mira variable sample} \label{sec:mirasample}

We use the O-rich Mira variable candidates selected in \citet{Zhang_2024}. We particularly choose O-rich Mira variables because they have a tighter period-luminosity relation \citep{Ita2011}. \citet{Zhang_2024} selected LPV with high \texttt{$\mathrm{best\_class\_score}$} and large $G$-band amplitude, $\Delta G > 0.4$, from the \textit{Gaia} LPV catalogue \citep{GaiaDR3_LPV} as Mira variable candidates. The O/C-rich classification is adopted from \citet{Sanders_Matsunaga_2023} when available. For those Mira candidates where classification is not available from \citet{Sanders_Matsunaga_2023}, we apply a series of color-period and amplitude-period cuts to remove C-rich Mira variable contamination. The luminosity distance of the O-rich Mira variables is assigned using the period-luminosity relation (PLR) in \citet{Sanders_2023}. The luminosity distance uncertainty is propagated from the intrinsic scatter of the PLR, the photometric uncertainty, and the uncertainty in period determination. We remove Mira candidates with periods shorter than $150$ days to avoid contamination from short-period(SP)-red stars, which removes Mira candidates older than $\sim11.2$~Gyr in terms of their ages. Furthermore, we exclude Mira candidates with distance modulus uncertainties greater than $0.6$ mag. In the final sample, there are 52 715 O-rich Mira candidates (see details of the sample construction and distance assignment in \citealt{ZS23} and \citealt{Zhang_2024}). The fractional distance uncertainty and period(age) distribution of the Mira variable sample are shown in the first and second panels of Fig.~\ref{fig:Mira_info}. The ages are assigned using the period-age relation for O-rich Mira variables in \citet{ZS23}. There are two period-age relations in \citet{ZS23}; one is from pure kinematic calibration, while the other averages over the kinematic ages of field Mira variables and the globular cluster ages for Mira variable globular cluster members. We adopt the latter one, although the difference between these two relations is not huge. \citet{ZS23} claimed a $\sim 11\%$ uncertainty in ages at a fixed period in the adopted period-age relation, but this uncertainty was derived by only accounting for the difference between the kinematically calibrated age and the Mira cluster members' isochronal ages. Hence, this reported age uncertainty is not robust and should only be treated as an order-of-magnitude estimation (see more discussion in \citealt{ZS23}). The median fractional distance uncertainty of the Mira variables is 12$\%$, similar to that of the RRL sample. The spatial $x$-$y$ and $x$-$z$ distributions of the O-rich Mira candidates are shown in the third and fourth panels. Most of the Mira variables reside close to the Galactic mid-plane ($|z|<3$~kpc), while many metal-poor RRLs are found farther from the plane. Similarly to RRL, we show the kinematic evolution of Mira variables as a function of pulsation period (age) in the top panel of Fig~\ref{fig:columnnormalised_vb}. The kinematics gradually becomes colder for longer-period Mira variables, which illustrates Mira's period-age relation. 

\subsection{Coordinate transformation}

For the RRL and Mira variables, only a small fraction of them have radial velocity measurements in {\it Gaia} DR3. This is because Mira variables have large pulsation amplitude that induces large systematic uncertainty in the line-of-sight velocity measured using a single exposure; while RRL suffers the same issue, the faint magnitude of RRL also makes the line-of-sight velocity measurement unreliable in {\it Gaia} DR3, especially for those in the disc plane with heavy dust extinction. Therefore, we limit our kinematic analysis to 5D phase space (three spatial coordinates and two on-sky velocities).

We adopted a left-handed Galactic frame or reference with the Sun located at $x_\odot = -8.122$ \citep{GRAVITY_2018}. We calculate the velocity of each star only in the Galactic longitude and latitude directions, correcting for the velocity of the local standard of rest (LSR) and solar reflex motion. We adopt the LSR velocity as $V_{LSR} = 238$~km/s \citep{Schonrich_2012} and the solar motion as ($U_\odot$, $V_\odot$, $W_\odot$) = (11.10, 12.24, 7.25)~km/s \citep{Schonrich_2010}. The velocities along the Galactic longitude, $v_\ell$, and latitude, $v_b$, are calculated as:
\begin{align*} 
v_\ell &= 4.74 \times \mu_\ell D + v_{\ell,\odot}, \\
v_b &= 4.74 \times \mu_b D + v_{b,\odot}, 
\end{align*}
where $\mu_\ell$ and $\mu_b$ are the proper motions in the Galactic longitude and latitude directions, respectively, in units of mas/yr, and $D$ represents the heliocentric distances in units of kpc. We then use $v_\ell$ and $v_b$ to analyse the kinematics of the RRL and Mira variables.

\section{Methodology}

\label{sec::methodology}

Our Galaxy has a well-established age-velocity dispersion relation, where older stellar populations exhibit hotter kinematics \citep{Stroemgren1946, Spitzer1951, Aumer2016, Frankel2020, Sharma_2021, Funakoshi2025}. Consequently, we can infer the ages of a stellar population from their kinematics. We use O-rich Mira variables as a natural clock due to their period-age relation \citep{Feast2000, Grady2019, TrabucchiMowlavi2022, ZS23}. By comparing the kinematics of RRLs with Mira variables at various periods, we can constrain the ages of RRLs under the assumption that two kinematically similar populations also have similar ages. In this section, we describe the methodology we have developed for a model-free comparison of the kinematics of two stellar populations (with only a 5D phase space, but it can be easily extended to 6D). We have verified our method using both idealised simulated galaxies (from Auriga cosmological simulation \citealt{Grand2017, Grand2024}) and our Galaxy, using a test sample of Mira variables. For convenience, we refer to the sample with unknown ages as \textit{sample A} (in the context of this paper,  RRLs in different metallicity bins), and the sample with known ages as \textit{sample B} (in our context, Mira variables with different periods).

\begin{figure*}
    \centering
    \includegraphics[width=1\linewidth]{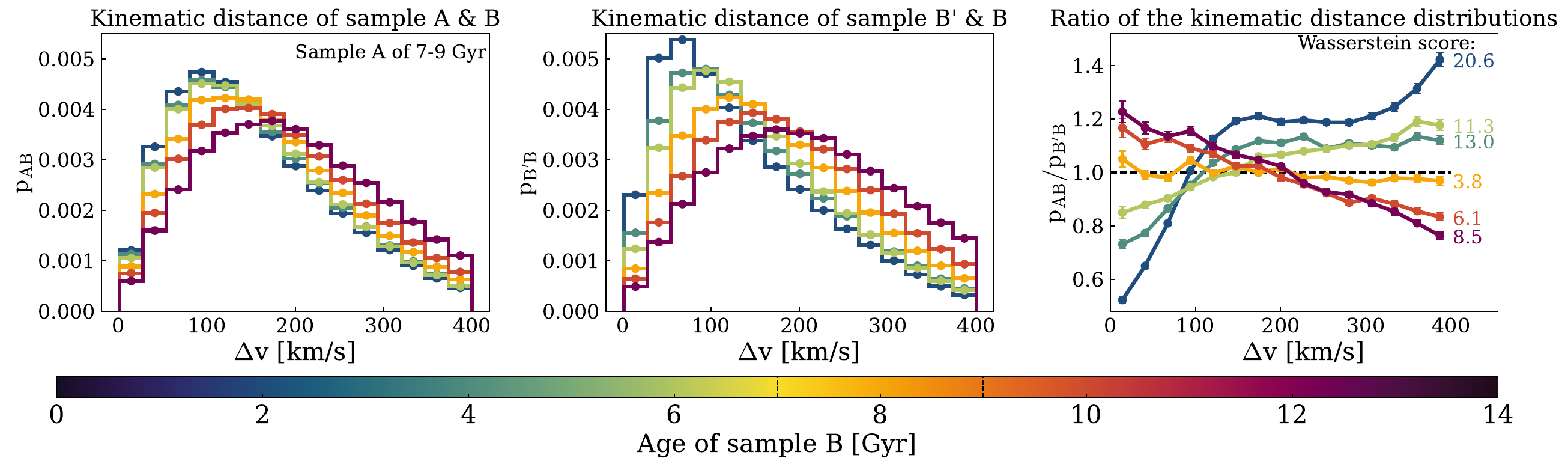}
    \caption{Demonstration of the methodology using an idealised simulated galaxy, Au18. {\it Left:} the velocity distances between sample A and B ,$p_\mathrm{AB}(\Delta v)$, approximated with a discrete histogram. {\it Middle:} Same as the top left but for $p_\mathrm{B'B}(\Delta v)$. {\it Right:} The ratio $p_\mathrm{AB}/p_\mathrm{B'B}(\Delta v)$, where the errorbar is propagated from the Poisson uncertainty for the approximated $p_\mathrm{AB}$ and $p_\mathrm{B'B}$. Sample A is fixed as a mono-age population sampled from particles in Au18 with ages between 7 and 9 Gyr. The results with different sample Bs of various ages are denoted in different colours. The Wasserstein scores (Wasserstein distances) of three example pairs are shown in the corresponding colour in the lower right corner. The ratio $p_\mathrm{AB}/p_\mathrm{B'B}(\Delta v)$ is close to unity for all $\Delta v$ when the ages of samples A and B are the same, which has the lowest Wasserstein score.}
    \label{fig:testing_with_Au18}
\end{figure*}

\subsection{Evaluating the similarity in the velocity space}
\label{sec::method_description}
We focus the comparison of samples A and B on the on-sky velocity space ($v_\ell-v_b$), as the configuration space is often affected by the spatial selection function. With a known distribution function, $f(\mathbf{x}, \mathbf{v})$, this comparison can be achieved by calculating (see \citealt{ZS23}):
\begin{equation} 
p(v_\ell, v_b|\mathbf{x}) = \frac{\int dv_{\mathrm{los}} f(\mathbf{x}, \mathbf{v})}{\int d^3v f(\mathbf{x}, \mathbf{v})}, 
\end{equation}
However, since the distribution function of the sample is unknown in our case, we adopt a data-driven approach.

For the $i^{\mathrm{th}}$ star in sample A, we select stars in sample B that reside within a spatial sphere of radius $r$, 
centred on this star. We then calculate the velocity distances of the selected sample B stars from the $i^{\mathrm{th}}$ star in sample A in the $v_\ell$--$v_b$ space. Given that the radius $r$ is sufficiently small, the resulting distribution of the velocity distance approximates
\begin{equation} 
p(\Delta v_i)\approx p(\Delta v_i|\mathbf{x_i}), 
\label{eqn:local_approx}
\end{equation}
where $\Delta v_i = \sqrt{(v_{\ell,i}-v_{\ell,B})^2 + (v_{b,i}-v_{b,B})^2}$.
The choice of the cross-matching radius $r$ is crucial, as a small $r$ leads to a small cross-matching sample and hence a large Possion error, and a large $r$ jeopardises the approximation in Eq.~\ref{eqn:local_approx} and leads to a large systematic bias. Having experimented with several values of $r$ (among $0.2-1$~kpc), we find that $r=0.6-0.7$~kpc provides the best balance given the sample sizes of the RRL and Mira variables. Hence, we choose $r=0.7$~kpc for all subsequent analyses.
To avoid over-representation of any sample A star that resides in a crowded region, a weight equal to the reciprocal of the number of matched B stars is assigned to the velocity distances of the sample A star. If fewer than five stars from sample B are matched, the star in sample A is discarded. (This discards $\sim10-20\%$ of the selected metal-rich RRL sample during the later kinematic age analysis). Repeating this procedure for each star in sample A and combining the velocity distances with their corresponding weights, we compute the distribution of velocity distances, which serves as a discrete approximation of
\begin{equation} 
p_{\mathrm{AB}}(\Delta v) = \int d\mathbf{x}^3 p(\Delta v|\mathbf{x}) \rho_\mathrm{A}(\mathbf{x}), 
\label{eqn:p_ab}
\end{equation}
where $\rho_\mathrm{A}(\mathbf{x})$ is the observed density distribution of sample A. We provide a cartoon diagram illustrating this procedure in Fig.~\ref{fig:appendix:cartoon} in Appendix~\ref{Appendix::cartoon_illustration}.

The same procedure for approximating $p_{\mathrm{AB}}(\Delta v)$ is applied to sample B'. Sample B' is created from sample B to mimic the spatial distribution of sample A, i.e., $\rho_{\mathrm{B'}}(\mathbf{x}) \simeq \rho_{\mathrm{A}}(\mathbf{x})$, ensuring that sample B' has the same kinematics as sample B but with a similar spatial distribution to sample A.
Practically, we find the spatially closest star in sample B for each star in sample A and discard those sample B stars that failed to be the closest match to all stars in sample A.
We then follow the same steps to approximate $p_{\mathrm{B'B}}(\Delta v)$ for sample B'. Finally, we take the ratio of the approximated $p_{\mathrm{AB}}(\Delta v)$ and $p_{\mathrm{B'B}}(\Delta v)$ to normalise $p_{\mathrm{AB}}(\Delta v)$. If the kinematic properties of sample A are similar to those of sample B, the ratio $p_{\mathrm{AB}}/p_{\mathrm{B'B}}(\Delta v)$ would be approximately 1. Otherwise, the ratio would deviate from unity.

\subsection{Verification with Auriga simulation suite}\label{subsec:demonstration}

We use a simulated idealised analogue of the Milky Way galaxy from the Auriga cosmological simulation suite, Au18, to examine the method. We do not discuss the properties of the Auriga simulations and Au18 in detail here, as all we require from Au18 is a realistic age-velocity dispersion relation. Details on the Auriga cosmological simulations and Au18 can be found in \citet{Grand2017, Fattahi_2019, Grand2024}. 
The kinematics of stellar particles vary as a function of age, which aligns with the prerequisites of the method. Therefore, Au18 serves as a useful laboratory for testing the methodology.

\begin{figure*}
    \centering
    \includegraphics[width=0.99\linewidth]{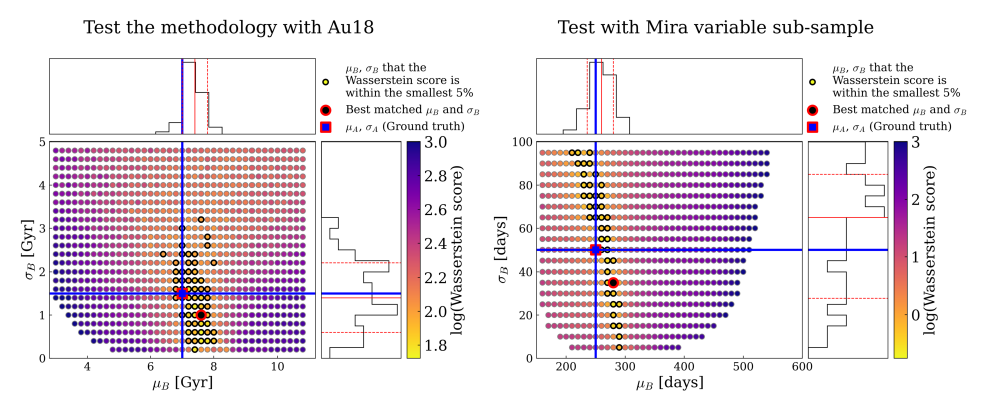}
    \caption{{\it Left panels: }Test result of the grid-search for modelling the age distribution of sample A from Au18. Each point on the grid represents a sample B constructed with an age distribution of $\mathcal{N(\tau|\mu_B,\,\sigma_B)}$ and is coloured with the logarithmic of the Wasserstein distances between the resulting $p_\mathrm{AB}(\Delta v)$ and $p_\mathrm{B'B}(\Delta v)$. The dots with black circles are those for which the Wasserstein distance is within the smallest $5\%$, and the histograms on the top and right show the $\mu_B$ and $\sigma_B$ distribution associated with these points. The red dashed and red solid lines on the distributions label the 16th, 84th and 50th percentile of the distribution. The black dot with a red circle is the grid point with the smallest Wasserstein distance, which denotes the best-fitted sample B so that ($\mu_B$, $\sigma_B$)$\approx$($\mu_A$, $\sigma_A$). The blue solid lines and the blue square with red edges denote the parameters for the Gaussian age distribution of sample A, i.e. $\mu_A$ and $\sigma_A$. {\it Right: }The same as the left set of the plots but for the test on the Mira variables, which verifies the method in the Milky Way environment. }
    \label{fig:Au18_Mira_Wassterstein}
\end{figure*}

We treat each stellar particle in Au18 as an individual star, although it actually represents a simple stellar population. We create sample A by selecting 5,000 particles with ages between 7 and 9 Gyr from Au18. Then, we bin the particles in Au18 by age and sample 50,000 particles from each age bin to create sample B, where the 5,000 particles in sample A are excluded to ensure that samples A and B are independent. 
We perform the analysis described in Section~\ref{sec::method_description} to compare the mono-age sample A with the sample B of various age bins. The results are shown in Fig.~\ref{fig:testing_with_Au18}. The left and middle panels show the distributions $p_{\mathrm{AB}}(\Delta v|\tau_A, \tau_B)$ and $p_{\mathrm{B'B}}(\Delta v|\tau_B)$, while the right panel shows the ratio $p_{\mathrm{AB}}/p_{\mathrm{B'B}}$, where $\tau_A$ and $\tau_B$ are the characteristic ages of samples A and B, respectively. Sample B from different age bins is denoted with different colours. We demonstrate that the ratio $p_{\mathrm{AB}}/p_{\mathrm{B'B}} \approx 1$ when $\tau_A = \tau_B$, as shown by the orange line. The ratio $p_{\mathrm{AB}}/p_{\mathrm{B'B}}$ exhibits a positive gradient when $\tau_A > \tau_B$. This is because sample B is kinematically colder than sample A, causing stars in sample B to cluster more in velocity space. Therefore, the distribution $p_{\mathrm{B'B}}(\Delta v|\tau_B)$ is more skewed towards smaller $\Delta v$ compared to $p_{\mathrm{AB}}(\Delta v|\tau_A, \tau_B)$. For the opposite reason, $p_{\mathrm{AB}}/p_{\mathrm{B'B}}$ shows a negative gradient when $\tau_A < \tau_B$. The greater the difference between $\tau_A$ and $\tau_B$, the more $p_{\mathrm{AB}}/p_{\mathrm{B'B}}$ deviates from unity.

After demonstrating that the method works in an idealised galaxy, we also examine how distance uncertainties affect the results, as such uncertainties can alter the kinematic properties of a stellar population. We repeat the above analysis, but this time, we manually redistribute the stellar particles in samples A and B by introducing a $10\%$ distance uncertainty. The results are shown and discussed in Appendix~\ref{Appendix::2PCF_witherr}. To summarise, we find that the method remains valid as long as the distance uncertainties in samples A and B are similar and sufficiently small, as for our comparison between RRL and Mira variables.

\subsection{Constraining the age distribution of sample A}


We also test a forward modelling approach to the method, in which we use sample B, with known ages, to constrain the unknown age distribution of sample A. This time, to better understand how sample size and differences in spatial distribution between the RRL and Mira samples impact the results, we ensure that sample A has a similar size and spatial distribution to the metal-rich RRL sample when sampling it from Au18, and sample B to have those characteristics similar to the Mira variable sample. As the ground truth, we select a subset of A with specific ages, by using a Gaussian window function parametrised by $\mu_A=7$~Gyr and $\sigma_A = 1.5$~Gyr. Therefore, the distribution function of sample A in phase space is
\begin{equation} 
f_A(\mathbf{x}, \mathbf{v}) = \int d\tau f(\mathbf{x}, \mathbf{v}|\tau)\mathcal{N}(\tau|\mu_A,\sigma_A), 
\end{equation}
where $f(\mathbf{x}, \mathbf{v}|\tau)$ is the extended distribution function conditioned on the stellar ages, $\tau$, for Au18, which we do not need to parametrise.

To infer the age distribution of sample A, we find $\mu_B$ and $\sigma_B$ of sample B such that the kinematic properties of sample B are identical to those of sample A using the methodology we developed. We adopt the Wasserstein distance (or Earth Mover's distance, or optimal transport distance, \citealt{Kantorovich1960, Vaserstein1969}) to quantify the difference between the velocity distance distributions $p_\mathrm{AB}$ and $p_\mathrm{B'B}$; a smaller Wasserstein distance would indicate greater similarity between $p_\mathrm{AB}$ and $p_\mathrm{B'B}$. We use the implementation in \textsc{scipy}\footnote{\url{https://docs.scipy.org/doc/scipy/reference/generated/scipy.stats.wasserstein_distance.html}} to calculate the optimal transport distance between the two 1D distributions $p_\mathrm{AB}$ and $p_\mathrm{B'B}$. For clarification, we refer to the Wasserstein distance as the Wasserstein score for the rest of the paper.  

We perform a grid search in the parameter space of $\mu_B$ and $\sigma_B$. For each grid point, we compile a sample B that follows a Gaussian age distribution, $\mathcal{N}(\tau|\mu_B,\sigma_B)$. We then repeat the previous steps of comparing samples A and B and calculate $p_\mathrm{AB}(\Delta v)$ and $p_\mathrm{B'B}(\Delta v)$. Next, we compute the Wasserstein score between the distributions $p_\mathrm{AB}$ and $p_\mathrm{B'B}$. The grid search results are shown in the left panel of Fig.~\ref{fig:Au18_Mira_Wassterstein}, where the colour represents the logarithm of the Wasserstein score corresponding to each grid point. The grid point with the smallest Wasserstein score is marked by a black dot with red edges, yielding ($\mu_B$, $\sigma_B$) = (7.6, 1.0)~Gyr. This is close to the ground truth mean and dispersion of the age distribution of sample A, which is ($\mu_A$, $\sigma_A$) = (7.0, 1.5)~Gyr and is indicated with blue lines and a blue square with red edges. We also select grid points where the associated Wasserstein scores are within the smallest $5\%$ and denote these points with black circles. The distribution of $\mu_B$ and $\sigma_B$ associated with these points is shown in the histograms on the top and right of the main panel. The 16th and 84th percentiles of the distribution are marked by red dashed lines, which provide a reasonable estimate of the uncertainty in our constraints. However, this is not exactly the uncertainty in a strict statistical sense. The 16th and 84th percentiles for $\mu_B$ and $\sigma_B$ are (7.0, 7.8)~Gyr and (0.6, 2.2)~Gyr, respectively. Both the best-fit values of $\mu_B$ and $\sigma_B$, as well as their 16th and 84th percentiles, are consistent with $\mu_A$ and $\sigma_A$. Therefore, we argue that our methodology can successfully constrain the age distribution of sample A in an idealized galaxy.

\subsection{Testing in the Milky Way environment}
\label{sec::test_with_mira}

We further validate the method in the Milky Way environment using the Mira variable sample. Similar to our previous analysis, we create a subsample from the Mira variable candidates with an underlying period (age) distribution that is treated as unknown for sample A. We then perform a grid search to optimise the parameters of sample B's period (age) distribution, aiming to compile a sample B that best fits the kinematics of sample A.

We first explore the case where the period distribution of sample A follows a single Gaussian distribution, $\mathcal{N}(\mathrm{period}|\mu_A, \sigma_A)$, with ($\mu_A$, $\sigma_A$) = (250, 50)~days. 
A grid search is performed to find the best-fit Mira variable subsample that has the closest kinematic properties. At each point on the grid ($\mu_B,\,\sigma_B$), we bin the Mira variables in period and randomly select Mira variables in each period bin so that the assembled Mira variable subsample has a period distribution close to a Gaussian distribution $\mathcal{N}(\mathrm{period}|\mu_B, \sigma_B)$. We draw 30 random realisations for each set of $(\mu_B,\,\sigma_B)$ and calculate the mean of the Wasserstein scores to ensure a fair comparison.
The grid search result is shown in the right panel of Fig.~\ref{fig:Au18_Mira_Wassterstein}. As in the left panel, the grid point with the smallest Wasserstein score is marked by a black dot with red edges, the points with the smallest $5\%$ of Wasserstein distances are labelled with black circles, and the ground truth values ($\mu_A$, $\sigma_A$) are shown as the blue lines and the blue square with red edges. The ($\mu_B$, $\sigma_B$) that gives the smallest Wasserstein score is (280, 35)~days, which is similar to the ground truth value. The histogram of $\mu_B$ for grid points marked with black circles also shows that the ground truth value of $\mu_A$ lies within the 16th and 84th percentiles. However, we find that the $\sigma_B$ distribution is quite broad, indicating that the constraints on the period dispersion could be weak given our sample size. Nevertheless, the result still provides a good constraint on the mean of the period (age) distribution, while the best-fit $\sigma_B$ is reasonably close to $\sigma_A$ as well.

 \begin{figure*}
    \centering
    \includegraphics[width = \linewidth]{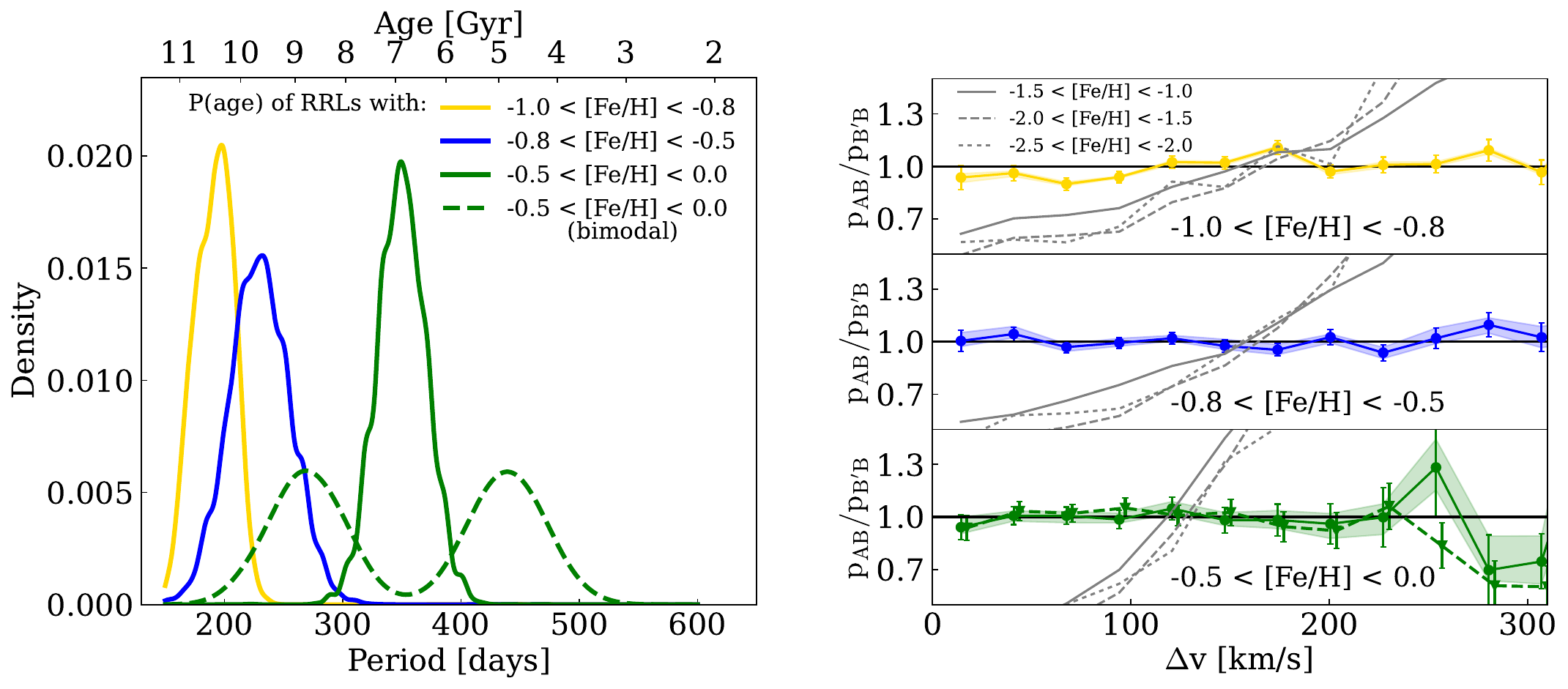}
    \caption{
    {\it Left panel: }the period (age) distribution of the Mira variable sample that has kinematics matched to RRL in different metallicity bins, which are equivalent to the age distribution of the RRL of the corresponding metallicity. The yellow histogram is for RRL with photometric metallicity between $-1<[\rm Fe/H]<-0.8$, blue for RRL with $-0.8<[\rm Fe/H]<-0.5$, and green solid for $-0.5<[\rm Fe/H]<0$. We also find that the kinematics of RRL with $-0.5<[\rm Fe/H]< 0$ are consistent with Mira variables with a bimodal age distribution, as shown in the green dashed histogram. {\it Right panels: } The coloured lines in each panel show the velocity distance ratios $p_{AB}/p_{B'B}$ of the RRL with different metallicities to their best-fit Mira variable sample, which shares the same legend as the left panel. The Possion uncertainty is presented in the error bar. The flatness of each of these lines demonstrates the goodness of fit. As a comparison, the grey solid, dashed, and dotted lines show the velocity distances ratio between the metal-poor RRL with metallicity of $-1.5<[\rm Fe/H]<-1$, $-2<[\rm Fe/H]<-1.5$, and $-2.5<[\rm Fe/H]<-2$ to the best-fitted Mira variable sample in the corresponding metallicity bins, e.g. {\it top right: }the velocity distance ratio between the metal-poor RRLs with Mira variables in the yellow histograms in the left panel. The positive gradient in all these lines demonstrates that RRL in these metal-rich bins is kinematically much colder than the metal-poor RRLs.}
    \label{fig:results_age_distribution}
\end{figure*}

\section{Results} \label{sec:results}

\subsection{Estimated age distributions of metal-rich RRLs}\label{subsec:age_given_met}

\begin{figure}
    \centering
    \includegraphics[width=\linewidth]{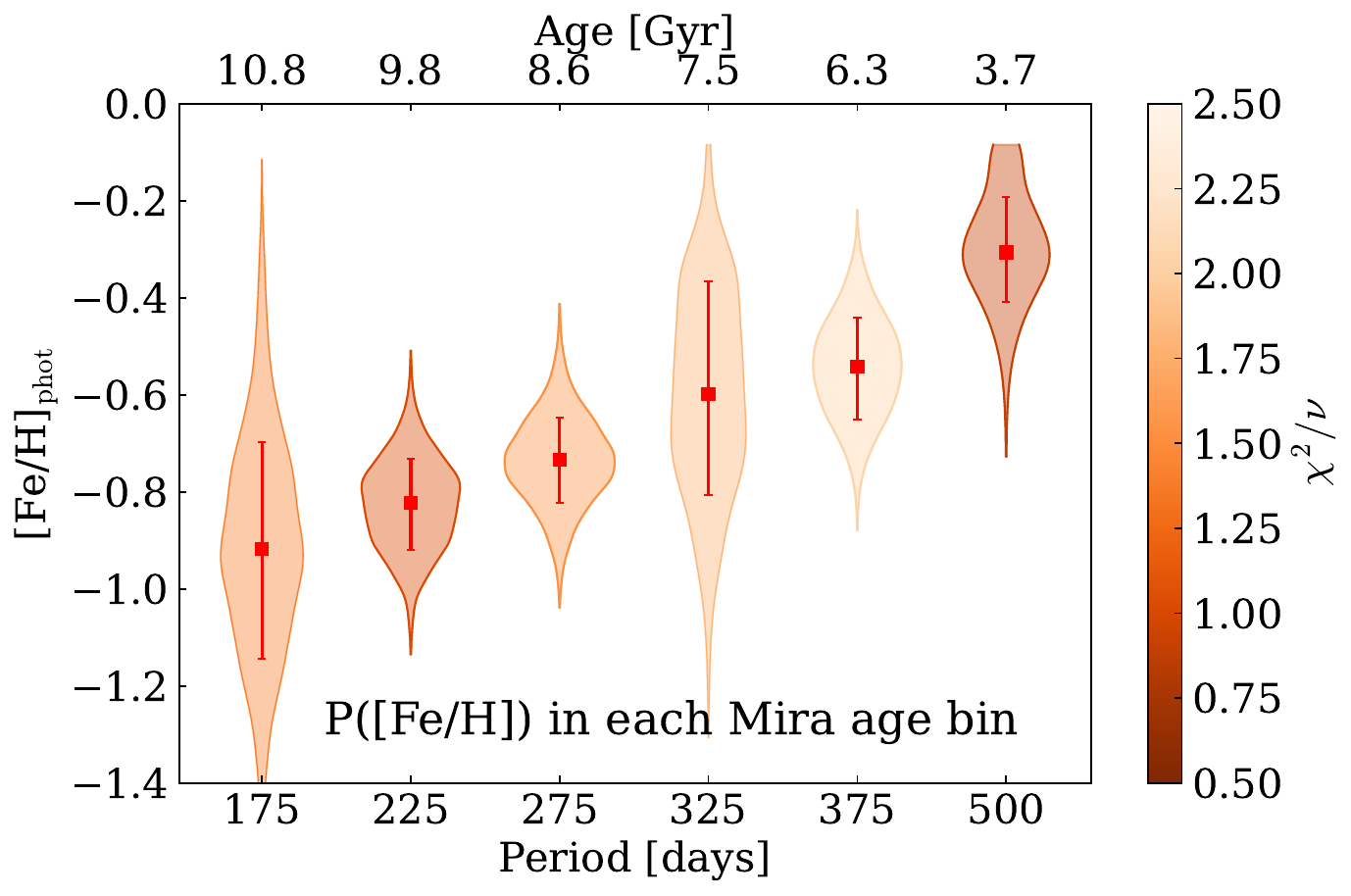}
    \caption{Best kinematically fitted metallicity distribution of the RRL sample for Mira variables in different period (age) bins. The colour represents the reduced $\chi^2$, $\chi^2/\nu$, for the velocity distance ratio compared to the unity. Younger Mira variables require a more metal-rich RRL population to have comparable kinematics. }
    \label{fig:P_feh_given_period}
\end{figure}

We apply the method described and tested in Section~\ref{sec::methodology} to the metal-rich RRL candidates and the Mira variable sample, in order to infer the age distribution of the metal-rich RRLs from their kinematics. As some of the metal-rich RRLs are discarded if they fail to find a sufficient neighbouring star in the Mira variable sample, so in total 2593 out of 3224 selected RRL with $\rm[Fe/H]_{phot}>-1$ are adopted for the analysis in this section. We bin the RRL candidates based on their photometric metallicity and use them as sample A in the context of the method. We compile a Mira variable subsample whose periods follow a specified period distribution as sample B. For each Mira variable subsample, we compute $p_\mathrm{AB}(\Delta v)$ and $p_\mathrm{B'B}(\Delta v)$ using the procedure outlined in Section~\ref{sec::method_description}. We minimise the Wasserstein score between $p_\mathrm{AB}$ and $p_\mathrm{B'B}$ so that the optimised Mira subsample has kinematics identical to the metal-rich RRL sample, implying that their age distributions are also similar due to the age-velocity dispersion relation in our Galaxy. We then use the period distribution of the optimised Mira subsample as a proxy for the ages of the metal-rich RRLs. To minimise the Wasserstein score, we repeat the analysis in Section~\ref{sec::test_with_mira}, performing a grid search to find the best-fit parameters for the period-interval selected subset of the Mira variables. 


In the left panel of Fig.~\ref{fig:results_age_distribution}, we show the period (age) distributions of the best-fit Mira variable subsamples, which serve as proxies for the underlying age distributions of RRLs in different metallicity bins.  The solid lines represent the results when assuming the age distribution is composed of a single Gaussian distribution. The panels on the right exhibit the ratio of $p_\mathrm{AB}/p_\mathrm{B'B}(\Delta v)$, with colours corresponding to the metallicity bins. The ratio $p_\mathrm{AB}/p_\mathrm{B'B}(\Delta v)$ for all coloured lines in each metallicity bin remains consistent with unity, implying that the kinematics of the RRLs in each bin are matched to those of the selected Mira subsample. 
The grey solid, dashed, and dotted lines show $p_\mathrm{AB}/p_\mathrm{B'B}(\Delta v)$ when comparing the kinematics of RRLs with metallicities of $-1.5\leq\mathrm{[Fe/H]}<-1$, $-2\leq\mathrm{[Fe/H]}<-1.5$, and $-2.5\leq\mathrm{[Fe/H]}<-2$, respectively, to the fitted Mira subsamples in the corresponding metallicity bins.

The yellow and blue lines suggest that the kinematics of RRLs with metallicities of $-1<\mathrm{[Fe/H]}<-0.8$ and $-0.8<\mathrm{[Fe/H]}<-0.5$ are consistent with the oldest Mira variables in our sample, which have periods distribution of $(\mu_B,\sigma_B)\sim(150-200, 5-50)$ days and $(\mu_B,\sigma_B)\sim(200-240,10-75)$ days, corresponding to ages of $\sim10.3-11.2$~Gyr and $\sim9.4-10.3$~Gyr \citep{ZS23}. (The reported range includes the grid search parameters with the Wassterstein score among the smallest 5$\%$). All the grey lines in the top two panels on the right have a positive slope, indicating that the kinematics of RRLs with $\mathrm{[Fe/H]}<-1$ are hotter than the oldest Mira variables in our sample. 

We also fitted the RRLs in these two metallicity bins with a bimodal Gaussian distribution in age. The resulting peaks of the age-bimodality are similar to each other and the age found for the single Gaussian fitting ($\mu_{1,B}\approx\mu_{2,B}\approx\mu_B$), so we do not further discuss the feasibility of bimodal age distribution here for RRLs of these two metallicity bins.


The solid green histograms in the left panel of Fig.~\ref{fig:results_age_distribution} suggest that the kinematics of RRLs with $\mathrm{[Fe/H]}>-0.5$ is consistent with Mira variables with a period distribution of a single Gaussian of $(\mu_B, \sigma_B)\sim(340-400,10-60)$~ days (corresponding to the ages of $\sim5.7-7.1$~ Gyr). We also find that their kinematics could also be consistent with Mira variables of a bimodal Gaussian with ($\mu_{1,B}, \,\mu_{2,B}\sim250-290,\,380-450$)~days (corresponding to ages of $\sim8.1-9.2$~Gyr and $\sim4.4-6.2$~Gyr, respectively). The Wasserstein score for both scenarios is similar, although the Wasserstein score for the single Gaussian fit is slightly lower, and the $p_\mathrm{AB}/p_\mathrm{B'B}$ ratios are both consistent with unity. This degeneracy is reasonable given the sample size of RRLs with $\mathrm{[Fe/H]}>-0.5$ is only around 250. A larger sample size in the future could help resolve this degeneracy. 

\subsection{Metallicity distribution of RRL at different ages} \label{subsec:met_given_age}

Adjusting the method slightly, we can supplement the results above by constraining the metallicity distribution of RRL at different ages.
To achieve this, we use all RRLs (with metallicities in the range $-2.5<\mathrm{[Fe/H]}<0$) as sample A and bin Mira variables in period segments as sample B. We then compute $p_\mathrm{AB}(\Delta v)$ and $p_\mathrm{B'B}(\Delta v)$ as before. In addition to the previous procedure, we assign an extra weighting factor to each star in sample A to reweight the distribution $p_\mathrm{AB}(\Delta v)$, based on the metallicities of RRLs in sample A, following $\mathcal{N}(\mathrm{[Fe/H]}|\mu,\sigma)$. We then optimise ($\mu, \sigma$) until the reweighted $p_\mathrm{AB}(\Delta v)$ and $p_\mathrm{B'B}(\Delta v)$ matches. This is equivalent to rewriting Eq.~(\ref{eqn:p_ab}) as
\begin{align*} 
p_{\mathrm{AB}}(\Delta v) = \int d\mathbf{x}^3 d\mathrm{[Fe/H]} & p(\Delta v|\mathbf{x},\mathrm{[Fe/H]})  \rho_\mathrm{A}(\mathbf{x}) \times \\
& \underbrace{\rho_A([\mathrm{Fe/H}]) \mathcal{N}([\mathrm{Fe/H}] | \mu, \sigma)}_{\mathrm{P([Fe/H])}},
\label{eqn:p_ab_2}
\end{align*}
where $\rho_\mathrm{A}(\mathrm{[Fe/H]})$ is the observed metallicity distribution of the RRLs, so $\mathrm{P([Fe/H])}$ is the weighted metallicity distribution of RRLs. When $p_\mathrm{AB}(\Delta v)$ and $p_\mathrm{B'B}(\Delta v)$ match, $\mathrm{P([Fe/H])}$ represents the optimised metallicity distribution of RRLs that fits the kinematics of Mira variables in a given period (age) bin. We minimise the difference between $p_\mathrm{AB}$ and $p_\mathrm{B'B}$ by minimising the Wasserstein score. The weighted metallicity distributions of RRLs in each of the age bins are shown in Fig.~\ref{fig:P_feh_given_period}. The square dots represent the mean of the weighted metallicity distributions, and the error bars denote the 16th and 84th percentiles. The results suggest that the kinematically colder and younger RRL populations are composed of RRLs with higher metallicities. These results are qualitatively consistent with the predictions in \citet{Bobrick2024} (see Section \ref{sec:implications}). The kinematics of the weighted RRL are in good agreement with the Mira variables in the correponding age group as demonstrated in Figure.~\ref{fig:met_given_age_kinematic_histogram}.


\section{Discussion} \label{sec:discussion}

\subsection{Comparison with other age estimates} \label{sec:comparisonage}

Since RRLs are evolved stars in the horizontal branch, a direct determination of their age (e.g., by comparison with stellar evolution models) is fundamentally unfeasible due to the strong degeneracy among age, mass loss along the red giant branch, and  chemical composition—particularly the  unknown helium abundance (see, e.g., \citealt{Marconi_2018,Savino_2020}). 
The age estimate therefore relies on comparisons with other stellar populations for which stellar ages are known (as in this work) and/or on empirical relations, such as the age–velocity dispersion relation (see Section \ref{sec::methodology}). In relatively simple and mono-age stellar populations, such as the one in stellar clusters\footnote{It is well known that Globular Clusters can host multiple stellar populations with variations in age and chemical composition \citep[see e.g.][]{Gratton_2019,Milone_2022,Gieles_2025}. However, for our discussion on the ages of RRLs in clusters, these differences are negligible, allowing us to assume each cluster as a simple, mono-age population.} or stellar streams, 
the age of member RRL is the same of the parent population that can be estimated with isochrone fitting or related methods \citep[see e.g.][]{VandenBerg_2013}.  Actually, the classification of RR Lyrae stars as old, metal-poor populations was originally based on their significant presence in globular clusters with metallicities in the range $-2 \lesssim \fehm \lesssim -1$ and ages exceeding $\approx 11$ Gyr, as well as in the Galactic stellar halo, which contains old ($\gtrsim 10$ Gyr) and metal-poor ($\fehm \lesssim -1$) relics of Galactic formation (see, e.g., \citealt{Kilic_2019,Horta_2024}). Based on this assumption, the presence of RRLs in a stellar population is often used as an independent indicator to confirm its old age (see e.g. \citealt{BSilva_2021,Feuillet_2022}).  
However, our results suggest that metal-rich RRLs are kinematically young, and can have ages down to a few Gyr.
It is therefore instructive to compare our findings with other age estimates for RRLs in the Galactic field and other environments.

\subsubsection{Metal-rich RRLs in the field: Disc} \label{sec:metrichdisc}
The first tentative connection between the kinematics of field RRLs and young stellar populations dates back to \cite{Strugnell_1986}. More than three decades later \cite{IB21}, leveraging the unprecedented RRL Gaia catalogue \citep{Clementini_2019}, quantitatively extended this analysis across the entire Galactic disc. By comparing the velocity dispersion of RRLs with the age-metallicity-angular momentum-velocity dispersion relations derived by \cite{Sharma_2021} for main-sequence and red giant branch disc populations, they demonstrated that the high-angular-momentum RRL population are kinematically cold enough to be consistent with young ages (2–5 Gyr, see also \citealt{Prudil_2020} for similar results in the solar neighbourhood). Despite possible caveats (e.g. assumption of axisymmetric, marginalisation over unknown line of sight velocity) their findings strongly suggest that these stars are not compatible with an age of $>10$ Gyr. Recently, \cite{Cabrera_2024_warp} conducted a similar study, relaxing the assumption of axisymmetry and finding that high-angular-momentum RRLs trace the warped structure of the Milky Way disc in a manner similar to red clump stars with ages around 3–5 Gyr (see \citealt{Das_2024}). Moreover, these RRLs exhibit a similarly low velocity dispersion.
In both \cite{IB21} and \cite{Cabrera_2024_warp}, RRLs were not selected based on metallicity; however, in both cases, their photometric metallicity distributions differ significantly from that of the bulk of the Galactic RRL population and are strongly skewed toward higher metallicities ($\fehm > -1$) and extended up to $\fehm > -0.5$. The age estimates of these works are qualitatively consistent with our results, especially considering the most metal-rich RR Lyrae in our sample (Figures \ref{fig:results_age_distribution} and \ref{fig:P_feh_given_period}). Independently of the absolute age estimates, which could be affected by several systematics, all these works strongly point toward a trend in which the metal-rich RRLs show kinematics similar to the intermediate age populations in the Milky Way disc.

It should be noted that, despite confirming the same kinematic trends, other authors do not necessarily agree on the connection between cold kinematics and young ages.
Indeed, while these stars exhibit overall $\alpha$-element abundances consistent with  the thin disc populations, their low abundance  of certain elements, such as yttrium, barium, scandium see \citep{,Gozha_2021,DOrazi_2024,Gozha_2024}, are more typical 
of older stellar populations (>10 Gyr). This chemical signature  is unexpected for genuinely intermediate-young disc populations, and appears to be in tension with standard nucleosynthesis models and Galactic chemical evolution theories \citep[e.g.][]{Horta_2022,Sheminova_2024}.  
Based on that results, \cite{DOrazi_2024} suggested that  metal-rich RR Lyrae stars may not belong to a conventional thin disc population.
Additionally, the authors stated that since both halo-like, low-angular-momentum and disc-like, high-angular-momentum stars are present in similar numbers in the solar neighbourhood, they must have followed a similar evolutionary pathway. 
Regarding the observed chemical peculiarities, we note that they may be explained by young stellar population (see Section \ref{sec:implications}).
When considering the relative numbers of halo-like to disc-like RR Lyrae stars, it is essential to account for the significant difference of about two orders of magnitude in the mass budget between the halo and the thin or thick disc (even when considering only the oldest disc components; see e.g. \citealt{Robin_2023}) as well as their different overall mass distributions.  
This translates into a difference in formation efficiency of at least 3 orders of magnitude (see e.g. Table 3 in  \citealt{Bobrick2024}). Although this information alone is not sufficient to confirm or rule out different formation scenarios, it must be considered when comparing the number of RR Lyrae stars at different metallicities.

Recent studies by   \cite{BSilva_2021}, \cite{Wu_2023} and \cite{Borbolato_2025}  suggest that the thin disc may have formed together with the thick disc and could therefore host old stars (>9--10 Gyr). Indeed, the most metal-poor RRLs in our sample are consistent with ages >10 Gyr and could represent a population of chemically defined thick disc stars, old thin disc stars, or a combination of both.  
Interestingly, \cite{Wu_2023}, using asteroseismic ages, dates the oldest thin disc stars in their sample to 9–10 Gyr, which is fully consistent with our results.

\subsubsection{Metal-rich RRLs in the field: Bulge area}

\cite{Savino_2020}, relying on single-star evolution models, conducted a comprehensive study of RRLs in the inner regions of our Galaxy, concluding that the bulk of these stars are extremely old, among the oldest in the Galaxy ($\approx 13$ Gyr).  
Despite most RRLs in this region being metal-poor ($\fehm < -1$), a tail of metal-rich objects (up to solar metallicity) exists. According to their population model, these metal-rich stars must be older than their metal-poor counterparts, with stars above $\fehm = -1$ inferred to be older than the Hubble time. The authors speculated that helium-enhanced populations could produce metal-rich RRLs with younger ages (see Section \ref{sec:implications}); however, for the most metal-rich RRLs ($\fehm > -0.5$), the required initial helium abundance would be significantly high ($Y > 0.3$). It remains to be assessed whether the required values are compatible with the extreme helium abundances inferred in some globular clusters (e.g., NGC 2808; \citealt{Milone_2015}).

Interestingly, a recent study by \cite{Prudil_2025} reports a clear distinction in the spatial distribution of RRLs in the bulge region based on their metallicity. Metal-rich RRLs ($\fehm > -1$) appear to trace the bar structure, whereas the metal-poor ones ($\fehm < -1$) show little or no evidence of a barred structure (see also \citealt{Olivares_2024, Zoccali_2024}). 
Due to kinematic fragmentation \citep{Fragkoudi2017, Debattista2017}, stars that were born before disc formation show halo-like kinematics in the bulge region, while populations born after disc formation show bar-like kinematics in the present day \citep{Zhang_2024}; the younger the population, the stronger the bar signature. Based on these findings, the authors speculate that the metal-rich sample must be younger than the metal-poor one. This conclusion contrasts with the population model proposed by \cite{Savino_2020} and could support the scenario in which metal-poor and metal-rich RRLs follow distinct formation and evolutionary paths (see Section \ref{sec:implications}).  
However, it is important to note that in a separate study, \cite{Han_2024} found no clear evidence of kinematic differences based on metallicity for the RRL population in the bulge region.

In conclusion, the origin of metal-rich RRL in the bulge area remains actively debated and no strong constraints on their age  have so far been derived. 
 
\subsubsection{Metal-rich RRLs in star clusters} \label{sec:clusters}

\begin{figure}
    \centering
    \includegraphics[trim=0cm 0cm 0cm 0cm, clip, width=\linewidth] {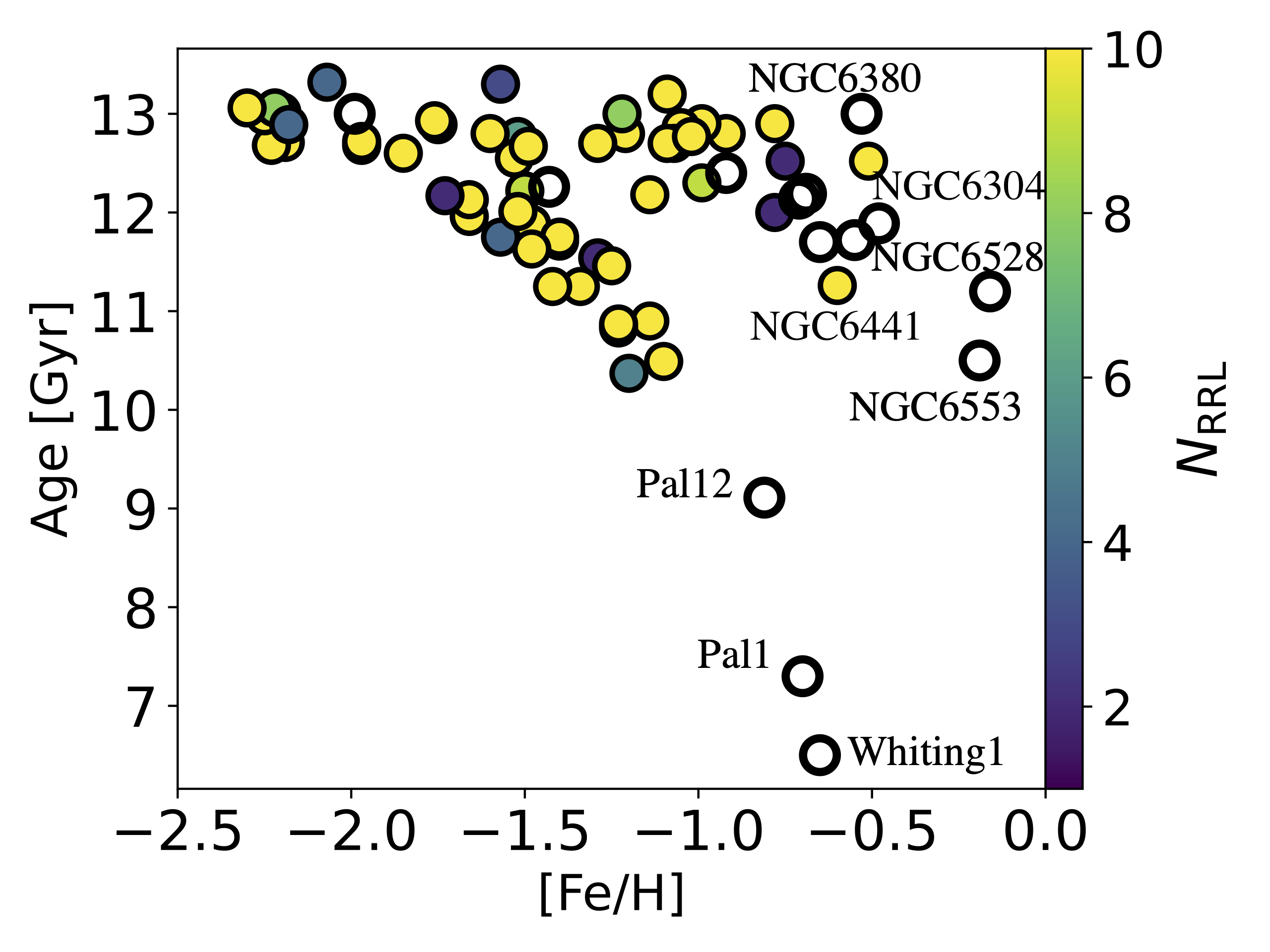}
    \caption{Number of RRL members in  75 Globular Clusters (colour map) as a function of metallicity and age of the clusters. Empty points indicate clusters devoid of RRLs. The data come from a crossmatching between the RR Lyrae membership from  \protect\cite{Cruz_2024} and \protect\cite{Alonso_2025} and the age and metallicity from \protect\cite{Kruijssen_2019} (56 clusters, average values in their Table A1). For   18  clusters  the ages and metallicities are  from  \protect\cite{Demarque_1992}, 
    \protect\cite{Feltzing_2002},
    \protect\cite{VandenBerg_2013}, 
    \protect\cite{Valcheva_2015}, 
    \protect\cite{Jahandar_2017}, 
    \protect\cite{Cohen_2021},  \protect\cite{Pallanca_2021}, and \protect\cite{Alonso_2025} (and reference therein). Formal uncertainties  on the age estimate from isochrone fitting are <0.5 Gyr, while systemic uncertainties due to distance and chemical abundance are $\approx1.5-2$ Gyr \protect\citep{VandenBerg_2013}. The data used  for the plot can be found in   \url{https://zenodo.org/records/15079962}.}
    \label{fig:NRRL}
\end{figure}

Figure \ref{fig:NRRL} shows that Milky Way globular clusters with a significant number of RRLs ($N_\mathrm{RRL}\gtrsim10$) predominantly span a metallicity range of $-2.5 \lesssim \fehm \lesssim -1$ and have ages between 11 and 13 Gyr. Notably, a marked decline in RRL counts is observed around $\fehm = -1$ (see also Figure 16 in \citealt{Cruz_2024}), which may indicate a drop in their formation efficiency. Indeed, most globular clusters at intermediate-to-high metallicities lack RR Lyrae, with only exceptions in the range $-1 \lesssim \fehm \lesssim -0.8$. In this metallicity interval, the RRLs in our sample exhibit kinematics similar to those of the oldest Mira stars in the disc. Our age estimate for these stars falls between 10 and 11 Gyr, somewhat lower than the approximately 12–13 Gyr typical of globular clusters; however, considering all potential uncertainties and systematic effects, the two estimates could marginally overlap.
At the highest metallicities ($\fehm \gtrapprox -0.5$), only two globular clusters in the analysed sample host RRLs: NGC 6441 ($\fehm > -0.6$, age $\approx$ 11-14 Gyr, \citealt{Kruijssen_2019,Massari_2023}) and NGC 6304 ($\fehm \approx -0.37$, age $\approx$ 11-14 Gyr, \citealt{Kruijssen_2019}). These clusters are significantly older than our field RRLs at similar metallicities. In addition, the RRLs in the cluster exhibit much longer periods compared to field RRL at similar metallicity.
NGC 6304 contains 11 RRLs, but the association of most of them is highly uncertain due to differential reddening \citep{DeLee_2006}. 
The two most likely members have long periods similar to those observed in NGC 6441.
Among the globular clusters not included in our sample, NGC 6388 shares a similar metallicity and a comparable blue extension of the horizontal branch with NGC 6441 \citep{Busso_2007}, and hosts 19 RRLs \citep{Cruz_2024}.
The presence of RR Lyrae stars in these three metal-rich and old clusters appears to be linked to a sub-population of helium-enriched horizontal branch stars \citep{Prtizl_2000, Caloi_2007, Sollima_2014}. However, \citet{Bhardwaj_2022b} found no evidence of helium enrichment in the near-infrared light curves of RRLs in NGC 6441.

Finally, it is interesting to discuss the cases of the relatively young globular clusters Pal12, Pal1, and Whiting1, and the most metal-rich clusters NGC 6528 and NGC 6553. Pal12 has an age of approximately 9 Gyr and metallicity $\fehm\approx-0.8$ \citep{Kruijssen_2019}, which falls within the range covered by field RRLs; however, no RRLs are found in this cluster. The age of Pal12 is uncertain—for example, \cite{Geisler_2007} reported an age of 6.5 Gyr—which, at such metallicity, would be too low even when considering our age estimates. 
Pal1 and Whiting1 are among the youngest globular clusters in the Milky Way (approximately <7 Gyr) with metallicities of $\fehm\approx-0.6$ to $-0.5$, and they are also devoid of RRLs \citep{Jahandar_2017,Valcheva_2015}. At these metallicities, we estimate that most field RRLs have ages greater than 7 Gyr. 
Both Pal 1 and Whiting 1 are among the least massive globular clusters (GCs), with masses below $2000$ M$_\odot$ \citep{Kruijssen_2019}. Assuming a similar RR Lyrae formation efficiency, we expect 2–3 orders of magnitude fewer RRLs compared to more massive clusters of similar metallicity (e.g., $\approx$10–30 RRLs in NGC 6304, NGC 6380, and NGC 6441), a result consistent with the absence of detected RRLs in these clusters. The clusters NGC 6528 and NGC 6553 have near-solar metallicities ($\gtrsim -0.2$), ages exceeding 10 Gyr, and lack RRLs  \citep{Demarque_1992, Feltzing_2002, Alonso_2025}. This absence, in combination with their high metallicity and old ages, is consistent with the properties of field RRLs.

We are not aware of any robust detections of RRLs in Milky Way open clusters. Most of these clusters are significantly younger (<1 Gyr) than the RRL ages estimated in this work \citep{Bossini_2019}. 
The oldest known open clusters \citep{ABaena_2024}—NGC 188 (age $\approx$ 7 Gyr, $\fehm \approx -0.03$), NGC 6819 (age $\approx$ 8 Gyr, $\fehm \approx -0.03$), and NGC 2682 (age $\approx$ 3.6 Gyr, $\fehm \approx 0.04$), and NGC 6791 (age $\approx$ 9 Gyr, $\fehm \approx 0.3$)—span an age and metallicity range that  partially overlap with field RRL. However, \citealt{Sandquist_2003} did not detect any RRL in NGC 2682, and only a few candidates ($\leq$3) have been found near NGC 188, NGC 6819, and NGC 6791, which were ultimately classified as non-members based on their magnitudes and proper motions \citep{deMarchi_2007,Sanjayan_2022b,Sanjayan_2022,Song_2023}.


Outside the Milky Way, \cite{Cuevas_2024} reported the likely association of 23 RRLs with Magellanic stellar clusters characterized by young-to-intermediate ages (1--8 Gyr) and metallicities ranging from $\fehm=-1.5$ up to solar values. This work independently corroborates the results of \cite{Sarbadhicary_2021}, who found indications that about half of the RRLs in the LMC could belong to young-to-intermediate populations (< 8 Gyr). The metallicity and age ranges of the RRLs associated with  Magellanic Clouds clusters are consistent with our findings. However, most of them are members of very young clusters (1--2 Gyr), apparently much younger than the field RRLs in our Galaxy. Among the others, four RRLs are associated with  the LMC cluster NGC 2121 (age  $\approx$ 3.2 Gyr), and the remaining are  found in the SMC clusters NGC 339 (age  $\approx$ 6 Gyr) and NGC 361 (age  $\approx$ 8 Gyr). While the ages  are consistent with those of disc RRLs, their metallicities ($\fehm\approx-0.6$ for NGC 2121, and $\fehm\approx-1.5$ for the others) are more typical of older Galactic RRLs.

In conclusion, the comparison between our age estimates for field RRLs and those derived from stellar clusters reveals partial overlap, along with some potential tensions.
However, it is crucial to consider potential systematic uncertainties that could affect the comparison, such as potential differences in the adopted age and metallicity scales  (see e.g. \citealt{VandenBerg_2013}, and Section \ref{sec:caveats}). For example, employing the steeper Mira age--period relation by \cite{Grady2019} (see Appendix \ref{Appendix:Different_ages}) would yield ages $\lesssim 2$ Gyr for the  the most metal-rich ($\fehm>-0.7$) RRLs, thereby bringing  them into better agreement with the ages and metallicities of RRLs observed in young clusters of the Magellanic Clouds \citep{Cuevas_2024}.

Additional sources of uncertainty include variations in detailed chemical compositions (e.g., $\alpha$-elements and helium), poorly constrained RRL formation efficiencies, and substantial differences in the stellar mass content of clusters and field populations. 
Specifically, globular clusters typically span the mass range $\sim10^4$–$10^5$ M$\odot$, young clusters in the Milky Way are less massive ($\sim10$–$10^3$ M$\odot$), while those in the Magellanic Clouds tend to be more massive ($\sim10^4$–$10^5$ M$\odot$; \citealt{Kruijssen_2019,Baumgardt_2013}). In contrast, the stellar mass of the Galactic disc field population is significantly larger, on the order of $\sim10^{10}$–$10^{11}$ M$\odot$ \citep[e.g.,][]{Robin_2023}. 
 As such, the formation efficiency of RRLs, and its dependence on age and metallicity, represents a critical parameter for comparing model predictions and observations across different environments.
Based on the detection of RRLs in young Magellanic clusters, \citet{Cuevas_2024} estimated a formation efficiency of $\sim10^{-5}$ M$_\odot^{-1}$ for intermediate-age RRLs, consistent with their absence in young clusters of the Milky Way. However, this value is approximately an order of magnitude higher than predictions from the binary evolution channel proposed for young/intermediate-age RRLs by \citet{Bobrick2024} (see Section \ref{sec:implications}).

Furthermore, the active dynamic environments facilitate stellar and binary interactions that can significantly alter the evolution of stars compared to the more quite field environment \citep{FusiPecci_1993, Pasquato_2013}.

A thorough comparison between model predictions and the properties of both field and cluster RR Lyrae stars, including a detailed assessment of systematic uncertainties, lies beyond the scope of the present study and will be the subject of future work.

\subsection{Implication for the RR Lyrae formation channels} \label{sec:implications}

\begin{figure}
    \centering
    \includegraphics[trim=0cm 1.6cm 0cm 0cm, clip, width=\linewidth] {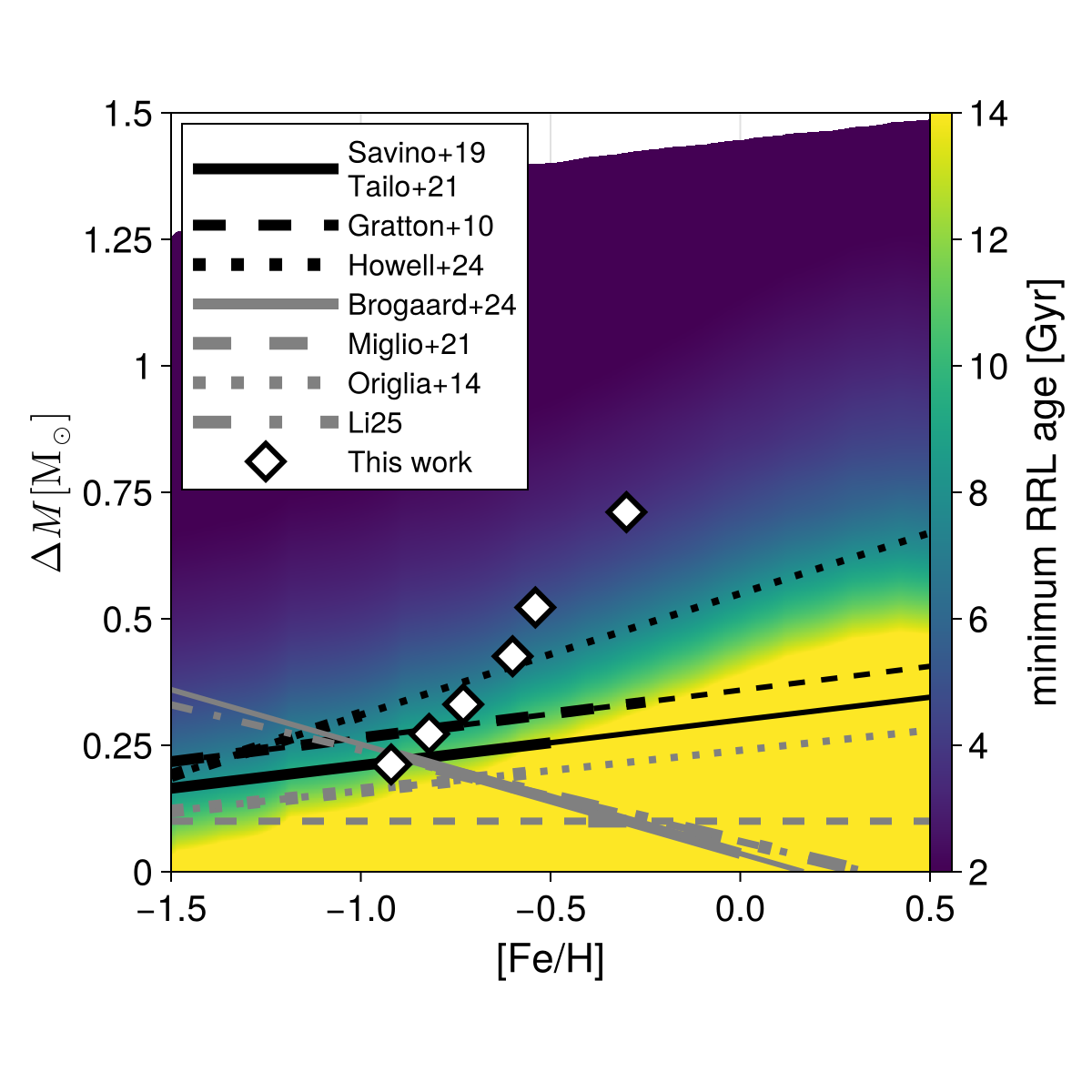}
    \caption{Minimum theoretical RRL age (colour map) as a function of metallicity and mass loss during the pre-core helium burning evolutionary stages, based on the \textsc{parsec} V1.2S stellar tracks and the instability strip boundaries from \protect\cite{Marconi_2015} (see Appendix \ref{app:parsecplot} for details). The different lines represent RGB mass-loss prescriptions calibrated on Globular Clusters \protect\citep{Gratton_2010, Origlia_2014, Tailo_2021, Howell_2024}, dwarf Spheroidals \protect\citep{Savino_2019}, and asteroseismology \protect\citep{Miglio_2021, Brogaard_2024, Li_2025}. The thicker segments of the lines indicate the metallicity range of the calibrators. White diamonds show the average age-metallicity trend from Fig.~\ref{fig:P_feh_given_period}, where the expected mass loss is self-consistently estimated from the metallicity–mass-loss–age relation shown in the colour map.}
    \label{fig:dmRRL}
\end{figure}

RRLs are formed from low-mass stars (with initial masses $<2$ M$_\odot$) that become hot enough to cross the instability strip during the core helium-burning phase \citep{Catelan_2015}, following the Helium Flash at the tip of the red giant branch (RGB) \citep{Catelan_2015}.
After the Helium Flash, these stars share a similar core mass ($M_\mathrm{c} \approx 0.5 \mathrm{M}_\odot$); thus, the temperature and the possibility to enter the instability strip depend on the metallicity and the amount of envelope mass lost (lower metallicity and less massive envelope produce higher temperature).
Given a specific metallicity and assuming the amount of mass lost in previous phases ($\Delta M$), one can determine the initial mass of the progenitor and hence the age of the RRL. Conversely, by knowing the age, as we have derived in this work, and the metallicity, we can infer the mass lost and thus discuss the possible sources of the physical mechanisms involved.

Figure \ref{fig:dmRRL} shows the RRL age predicted by the \textsc{parsec} V1.2S stellar evolution model \citep{Chen_2015} at a given metallicity (assuming $Z_\odot=0.0152$, with $\fehm = \log Z - \log Z_\odot$) and for a specific amount of mass lost prior to the formation of RRLs (additional details are in Appendix \ref{app:parsecplot}), together with wind mass loss estimates (black and gray lines), and results from our analysis (white diamonds). 
Below $\fehm \lesssim -0.8$, the age we derived for field RRLs implies relatively low mass loss ($\Delta M < 0.3$ M$_\odot$), which is consistent with some of the wind mass mass loss prescriptions calibrated on metal-poor environments \citep{Gratton_2010, Savino_2019, Tailo_2021}. At higher metallicities, the required mass loss rapidly increases from $\Delta M \approx 0.4 \ \mathrm{M}_\odot$ at $\fehm \approx -0.5$ to $\Delta M \gtrapprox 0.7 \ \mathrm{M}_\odot$ for $\fehm \gtrapprox -0.3$. This high value of mass loss significantly exceeds what is expected from winds, even considering the highest values extrapolated from the \cite{Howell_2024} relations (calibrated on $\fehm \lesssim -1$). In general, at high metallicity, the predicted wind mass loss is highly uncertain, but most relations predict $\Delta M \lessapprox 0.3 \ \mathrm{M}_\odot$ and therefore imply extremely long RRL ages ($>$13--14 Gyr), which are not consistent with our results and with the existence of metal-rich RRLs.
In particular, asteroseismological studies  at $\fehm > -0.5$ seems to find a very low or even decreasing wind mass loss efficiency as function of metallicity \citep{Miglio_2021,Brogaard_2024,Li_2025}.

These results suggest that the most metal-rich RRLs in the field are not a standard outcome of stellar evolution. Rather, as already noted by \cite{Tam_1976} and \cite{Bono_97a}, they are likely related to exceptional and extreme events of mass loss. The discovery of a peculiarly low-mass ($\approx$ 0.26 M$_\odot$) RRL-like pulsator \citep{Pietrzynski_2012, Smolec_2013} demonstrated that this exceptional source of mass loss can be ascribed to mass transfer in binary systems \citep{Karczmarek_2017}. \cite{Bobrick2024} conducted detailed population studies of RRL produced through binary mass transfer using binary evolution \textsc{MESA} models \citep{Paxton_2019} and realistic Galactic populations \citep{Robin_2023}. In their model, all the metal-rich RRLs ($\fehm>-1$) are produced by binary interactions and belong to the intermediate-young disc populations with ages ranging from 1 to 10 Gyr. Interestingly, the age-metallicity distribution shows a bimodal pattern, with ages within 8--10 Gyr for $-1 \lesssim \fehm \lesssim 0$, and 1--5 Gyr for $\fehm>-0.4$ (see their Figure 6). Although the absolute values of ages and metallicities show a quantitative offset (see Section \ref{sec:caveats}), the bimodality is qualitatively consistent with our results when a bimodal age distribution is assumed for the most metal-rich RRLs (see Figure \ref{fig:P_feh_given_period}). \cite{Bobrick2024} estimated a low formation efficiency for metal-rich RRLs ($\lesssim 10^{-6}$ M$^{-1}_\odot$), about three order of magnitude lower with respect to metal-poor RRLs formed through the conventional channel. This low rate is consistent with the observed number of metal-rich RRLs in the field, and with their paucity  in the Milky Way clusters (see Section \ref{sec:clusters}).
The rate is instead at least one order of magnitude lower than what inferred by the association of RRLs with young Magellanic Clouds in \cite{Cabrera_2024}.

Metal-rich RRLs, exhibit a peculiar combination of low $\alpha$-elements abundances and low abundances on elements such as yttrium, barium and scandium \citep[][see discussion in Section \ref{sec:metrichdisc}]{Gozha_2021, DOrazi_2024}.
This characteristic is not necessarily contradictory to a binary origin; rather, it mirrors anomalies observed in a class of heavily stripped stars—such as post-RGB and post-AGB stars \citep{vanWinckel_2003, Oudmaijer_2022, Mohorian_2024,Mohorian_2025}—which are  commonly associated with mass transfer episodes \citep[see, e.g.,][]{Gezer_2015, Kamath_2019, Oomen_2019, GallardoCava_2022}. 
A binary origin for metal-rich RR Lyrae stars has the potential to simultaneously resolve the tension between their kinematically inferred intermediate-young ages and their peculiar chemical compositions, which differ from those of the typical Galactic disk population \citep{Bensby_2014}.
However, while there are candidate RRLs in binary systems, robust identifications remain elusive, and the current observational data are insufficient to either confirm or definitively rule out the binary formation channel (see e.g. \citealt{Abdollahi_2025}, and Appendix A in \citealt{Bobrick2024}).

Other authors interpret the chemical peculiarity of metal rich RRLs as the signature of the non-native disc origin of such populations \citep[see e.g.][]{DOrazi_2024, Feuillet_2022}. In this case the RRLs could be genuinely very old and accreted in the Milky Way or migrated from the innermost part of our Galaxy. 
However, Figure \ref{fig:dmRRL} shows that, independently of their origin, at highest metallicity at least 0.4 M$_\odot$ must have been lost during pre-RRL phases. 
This amount of mass loss falls within the limits predicted by \cite{Howell_2024}, although these predictions are extrapolated from a relationship calibrated at much lower metallicities ($\mathrm{[Fe/H]} < -1$). If high wind mass loss is a common characteristic of low-mass, metal-rich stars, it could have significant implications for a wide range of other stellar populations, including the entire family of horizontal branch stars, red clump stars, and the initial-final mass relation of white dwarfs. Alternative sources of high mass loss rates may include eruptive events driven by rapid stellar rotation and strong magnetic fields \citep{Vidotto_2011, Ramsay_2020}. However, no specific studies have yet addressed the production of unusual hot core helium-burning stars resulting from these processes.

A possible alternative to high mass-loss rates for the formation of metal-rich RRLs is helium enrichment. Indeed, helium-enriched envelopes can produce hotter core-helium burning stars, counterbalancing the higher opacity of metal-rich stars. 
At a fixed effective temperature (i.e. within the instability strip), helium-enriched RR Lyrae stars are more luminous and have larger radii, resulting in lower mean stellar densities and hence longer pulsation periods \citep{Salaris_2005,Lee_2016}.
Notably, NGC 6441 is known for hosting helium-enriched stellar populations, and the associated RRLs have much longer periods than field RRLs at similar metallicities (see Section \ref{sec:comparisonage}).
\cite{Savino_2020} estimated that helium enrichment, up to $Y=0.3$,  is insufficient to produce very metal-rich RRLs ($\fehm \gtrsim - 0.5$) with ages under 14 Gyr. However, higher values (up to $Y\approx0.4$) are found in massive globular clusters (see e.g. \citealt{Milone_2015}).  Whether helium abundances greater than $Y = 0.4$ are sufficient to produce metal-rich RRLs consistent with the ones in the field remains to be investigated. 

\subsection{Metal-poor RRLs populations in the Disc region}

Our analysis in Section~\ref{subsec:age_given_met} highlights that the RRLs  with $\mathrm{[Fe/H]} < -1$ are kinematically hotter than the oldest Mira variables in our sample (see right panel of Figure~\ref{fig:results_age_distribution}). At face value, this would imply that such metal-poor RRLs in the disc are older than $11~\mathrm{Gyr}$ (Section~\ref{sec:mirasample}), representing a very old disc component consistent with recent findings \citep{BSilva_2021,Wu_2023,Borbolato_2025}. However, it is important to note that the old disc populations identified in those works are generally more metal-rich, and that inferring ages for metal-poor RRLs by extrapolating the age--velocity relation beyond the parameter space covered by the Mira sample is unreliable and prone to biases.

A more realistic interpretation is that such metal-poor RRLs in the disc are kinematically hotter than the Mira sample because they belong to a different kinematic component of the Milky Way. Their hotter orbits imply a broader spatial distribution, allowing them to occupy a larger volume of the Galaxy, including the innermost regions of the disc. Indeed, in the solar neighbourhood, the vast majority of RRLs with $\mathrm{[Fe/H]} < -1$ are associated either with the thick disc or the stellar halo \citep{Prudil_2020,Zinn_2020,DOrazi_2024}. This trend has been confirmed across the Galactic disc through the chemo-kinematic analysis of Gaia RRLs by \citet{IB21}. By crossmatching our sample with their classification \citep{IB21_zenodo}, we find that only 7\% of the RRLs with $\mathrm{[Fe/H]} < -1$ are associated with the fast-rotating disc component (broadly corresponding to the thin disc), while the remaining stars are classified as members of the stellar halo (67\%) or as part of a mixture of the slow-rotating disc (i.e., the thick disc) and halo substructures (26\%; see their Figure~15).The metal-poor stars associated with the fast-rotating disc may either be contaminants, due to uncertainties in the photometric metallicity classification, or genuine members of the halo or thick disc populations that happen to follow relatively cold orbits still consistent with the broad velocity dispersion characterising these components \citep[see e.g.][]{DOrazi_2025}.

We conclude that the metal-poor population of RRLs ($\mathrm{[Fe/H]} < -1$) analysed in our sample appears kinematically hotter than the Mira population due to its association with the stellar halo and thick disc. While our method cannot be directly applied to infer the ages of these stars, their low metallicities and connection to old Milky Way components strongly suggest that they belong to the classical family of RR Lyrae stars, with typical ages exceeding $10~\mathrm{Gyr}$.

\subsection{Caveats} \label{sec:caveats}

There are several limitations with our analysis that could be improved in the future.

\emph{Measurement uncertainty and bias: }As our methodology performs better with a larger sample size, we adopt the photometric metallicity, instead of the spectroscopic metallicity, of RRL in this work. The uncertainty on the photometric uncertainty is typically of the order $0.15-0.35~\rm dex$. Although we choose the bin width of the metal-rich RRL in Section~\ref{subsec:age_given_met} to be comparable to the scale of metallicity error, we should still expect metal-poor RRL contamination in these metal-rich bins. As we show in Fig.~\ref{fig:columnnormalised_vb}, metal-poor RRL are kinematically hotter than the metal-rich ones, presumably because they are constituents of the Galactic halo, the contamination from metal-poor RRLs leads to an overestimation on the velocity dispersion of the metal-rich RRLs. Therefore, the large uncertainty on the photometric metallicity likely leads to an overstimation of the ages of metal-rich RRLs. 

The quality of the photometric metallicity also affects the heliocentric distance measurement of RRLs as they are strongly correlated. The photometric metallicity of the metal-rich RRL tends to be underestimated \citep[see e.g.][]{IB21,Mullen_2021,Muraveva_2025}, which would cause their absolute magnitude to be underestimated and the heliocentric distance to be overstimated. This subsequently leads to an overestimation of their velocity dispersions as the velocity scales as $\mathbf{\mu}\times D$, and hence their age estimation tends to be older than they are, because of the age-velocity dispersion relation of the Milky Way. Overall, we conclude that the age of the metal-rich RRLs could be overestimated due to the accuracy and precision of the adopted photometric metallicity.

\emph{Period-age relation of Mira variables: }We used the Mira variable as a standard(isable) clock to determine the age of the RRLs. The period-age relation of the Mira variables we adopted \citep{ZS23} are calibrated based on their kinematics, or more specifically, the velocity dispersion, rather than measured from stellar models. Stars with more robust age measurement are available, such as isochronal ages of main-sequence turn-off (MSTO) stars \citep[e.g.][]{Xiang_Rix2022}, asteroseismic ages \citep[e.g.][]{Miglio_2021}. However, the isochronal and asteroseismic age measurements are mainly available for stars in the solar vicinity because of the applicability of methodology and the survey coverage, while the majority of metal-rich RRLs in our sample are in the inner Milky Way as shown in Fig.~\ref{fig:RR-Lyrae_info}. As our methodology requires a common spatial coverage for the RRLs and the comparison sample, it would further decrease the sample size of the metal-rich RRLs by $\sim95\%$ if the comparison sample cannot reach the inner Galaxy. In contrast, Mira variables are more abundant outside the inner Galaxy, allowing a more robust dynamical modelling, and hence, a better kinematic age calibration. Therefore, we use Mira variables as a comparison sample as it is abundant both in the inner and outer disc. Spectroscopic ages for red giant stars (e.g. \citealt{Leung2023} and \citealt{Anders2023}) have larger coverage towards the inner Milky Way. We repeat our approach in Section~\ref{subsec:met_given_age} but use the red giant stars in \citet{Anders2023} as a comparison sample. The results are broadly consistent with the results presented in Section~\ref{subsec:met_given_age}, strengthening our analysis using Mira variables. More details can be found in Appendix~\ref{Appendix:Different_ages}.

Mira variables have been demonstrated as a good age indicator in many observations (\citealt{Wyatt1983, Clement2001, Grady2019, ZS23}, see also \citealt{Feast2009, FeastWhitelock2014})
The period-age relation we adopted in this work~\citep{ZS23} is calibrated based on the age-velocicty dispersion relation in~\citet{YuLiu2018}, which are derived using isochronal ages. As shown in Fig.~13 of \citet{ZS23}, the kinematically calibrated period-age relation is compatible with the past results~\citep{Wyatt1983, Feast2009, FeastWhitelock2014} and Mira variable globular cluster members~(\citealt{Clement2001}, although the globular cluster Mira variables suggest that the Mira's ages to be $\sim$1~Gyr older than that in the period-age relation for period $\sim$250~days). It also agrees nicely with the theoretical results in \citet{Eggen1998}. The period-age relation calibrated using Mira variables in LMC clusters in \citet{Grady2019} is in apparent disagreement with the adopted relation, but the calibration in \citet{Grady2019} may be subject to contamination from the Mira variables of the LMC field. The recent theoretical modelling in \citet{TrabucchiMowlavi2022} suggests that LPV in the fundamental pulsation mode exhibits a large age spread in a fixed period, and the resulting theoretical period-age relation is consistent with that in \citet{Grady2019}. However, the large spread of the period-age relation in \citet{TrabucchiMowlavi2022} also makes it almost consistent with the relation adopted in this work. The foundation of the kinematic age calibration in this work would break if the age spread at a fixed period of Mira variables were too large. The adopted period-age relation in \citet{ZS23} has also been used to determine the time of disc formation and the time of bar formation \citep{Sanders2024, Zhang_2024} and has shown agreement with other studies using different age indicators \citep{Belokurov_Kravtsov2024, Haywood2024}. Therefore, we argue that the adopted period-age relation is still reliable, but a future calibration is necessary to confirm Mira's age spread at a fixed period. Also, all the modelling in this work is done in the period space of Mira variables, so the results could be easily revised when a better-calibrated Mira period-age relation arrives. The resulting age-metallicity correlation for adopting different Mira's period-age relation is included in Appendix~\ref{Appendix:Different_ages}. 

In summary, adopting other period-age relations in literature would not affect the main conclusion of the paper that the metal-rich RRL are much younger than the metal-poor ones. However, the age of RRL with metallicity of $\rm{-1<[Fe/H]}<-0.5$ could be older (age~$>10.5$~Gyr) than the value reported in this paper if we only used the ages of the Mira variables in globular clusters \citep{VandenBerg_2013}, but the period-age relation of Mira in globular clusters is less likely to be the same as it in field stars.

The period of Mira variable also appears to be correlated to the metallicity \citep{FrogelWhitelock1998, FeastWhitelock2000}. This correlation could be partially attributed to the age-metallicity relation of the globular clusters \citep{Belokurov_Kravtsov2024}, but we call attention to the fact that the metallicity dependence of the Mira variables' period-age relation has not yet been quantified. Due to the negative radial metallicity gradient along the Galactic disc \citep{Braganca2019}, the period-age relations of Mira variables on the inner disc and the outer disc could potentially be different. To understand this systematics, we divide the RRL and Mira variable samples into inner ($R<5$~kpc) and outer ($R>5$~kpc) disc regions, and repeat the analysis in section~\ref{subsec:met_given_age} on them, respectively. The results are shown in Fig.~\ref{fig:inner_outer_fit}. We see a systematic difference between the inner and outer disc fit, which could be attributed to the metallicity difference of the inner and outer disc stars, but the difference is very minor ($\delta[\mathrm{Fe/H}]_{\rm phot}\lesssim0.2$) in terms of the best-fit RRL metallicity in each period bins. Therefore, we argue that the metallicity dependence of Mira variables does not affect our conclusion much.

\section{Summary} \label{sec:conclusions}

In this work, we estimate the age of field metal-rich ($\fehm > -1$) RR Lyrae stars by comparing their kinematics (based on Gaia DR3 astrometry) with a sample of O-rich Mira variables, whose ages are determined through a period-age relation calibrated on the velocity dispersion-age relation for disc populations. After validating our method using the Auriga simulation suite, we obtain the following main results.

\begin{itemize}
\item Metal-rich RR Lyrae stars exhibit cold kinematics comparable to those of Mira variables in the disc. Conversely, metal-poor RR Lyrae stars consistently show hotter kinematics than those of the Mira populations, typical of a halo-like population. 
\item Metal-rich RR Lyrae age span a range typical of the intermediate-old population in the stellar disc, ranging from 3 to 11 years, challenging the traditional view of these stars as exclusively ancient (>10 Gyr).
\item We observe an age-metallicity trend among RR Lyrae stars. The most metal-poor RR Lyrae stars ($-1 \lesssim \fehm \lesssim -0.8$) exhibit kinematic similarities with the oldest Mira stars in the disc (10--11 years). At intermediate metallicities ($-0.8 \lesssim \fehm \lesssim -0.5$), the RR Lyrae stars are compatible with Mira populations with ages within 7--11 years. At the highest metallicity ($\fehm \gtrsim -0.5$), the RR Lyrae stars are younger, with ages ranging from 3 to 8 Gyr. 
\item For $\fehm \gtrsim -0.5$, we find evidence of a bimodal age distribution, with the older RR Lyrae peaking at 9 Gyr and the younger at 4--5 Gyr. 
\end{itemize}

Although the absolute value of our age estimate could be subject to systematics (e.g. on the Mira period-age calibration), the decreasing age-metallicity trend is based on a relative comparison and is a robust result of our analysis. 
The estimate of disc-like kinematics linked to intermediate age for the metal-rich RR Lyrae stars is consistent with other recent investigations on field RR Lyrae stars \citep{IB21,Cabrera_2024_warp}. The assessment of potential tensions between the RR Lyrae age estimates in the Milky Way, and Magellanic Cloud clusters require more specific and detailed studies.

The intermediate ages for the most metal-rich RR Lyrae stars, along with the inferred age-metallicity trend, is inconsistent with or even opposite to the predictions of the standard RR Lyrae formation channel, unless an exceptionally high amount of mass loss for the RR Lyrae progenitors is assumed. Among other formation channels, mass transfer in binary systems, which aligns with much of the observational evidence, emerges as a promising alternative \citep{Karczmarek_2017, Bobrick2024}. However, a conclusive identification of RR Lyrae stars in binary systems is still lacking \citep[see e.g.,][]{Abdollahi_2025}.

Future observations and further advancements in modelling are crucial for constraining the properties and origins of metal-rich RR Lyrae stars. Characterising this intriguing population is essential not only for revising our understanding of RR Lyrae  formation and evolution but  could also have significant implications for theories of single and binary stellar evolution, as well as for models of Galactic formation and evolution.


\section*{Acknowledgements}
We thank Cecilia Mateu, Alessandro Savino, and Michele Trabucchi for their helpful comments on the draft of the paper. We further thank Friedrich Anders and Marcin Bartosz Semczuk for suggesting the tests in Appendix~\ref{Appendix:Different_ages}.

HZ thanks the Science and Technology Facilities Council (STFC) for a PhD studentship. 
GI was supported by a fellowship grant from the la Caixa Foundation (ID 100010434). The fellowship code is LCF/BQ/PI24/12040020.
VB acknowledges support from the Leverhulme Research Project Grant RPG-2021-205: "The Faint Universe Made Visible with Machine Learning". A.B.~acknowledges support from the Australian Research Council (ARC) Centre of Excellence for Gravitational Wave Discovery (OzGrav), through project number CE230100016. VD acknowledges financial support from the Fulbright Visiting Research Scholar program 2024-2025 and PRIN MUR 2022 (code 2022YP5ACE) funded by
the European Union – NextGenerationEU.

\section*{Data availability}

\textit{Gaia} data used in this work is publicly available. The kinematic comparison code developed in this work is available at github\footnote{https://github.com/Hanyuan0908/kinematic\_label\_transfer}. Please contact the corresponding author for any questions.




\bibliographystyle{mnras}
\bibliography{bibliography} 

\appendix

\section{Illustration of the kinematics distance}
\label{Appendix::cartoon_illustration}

\begin{figure*}
    \centering
    \includegraphics[width = \linewidth]{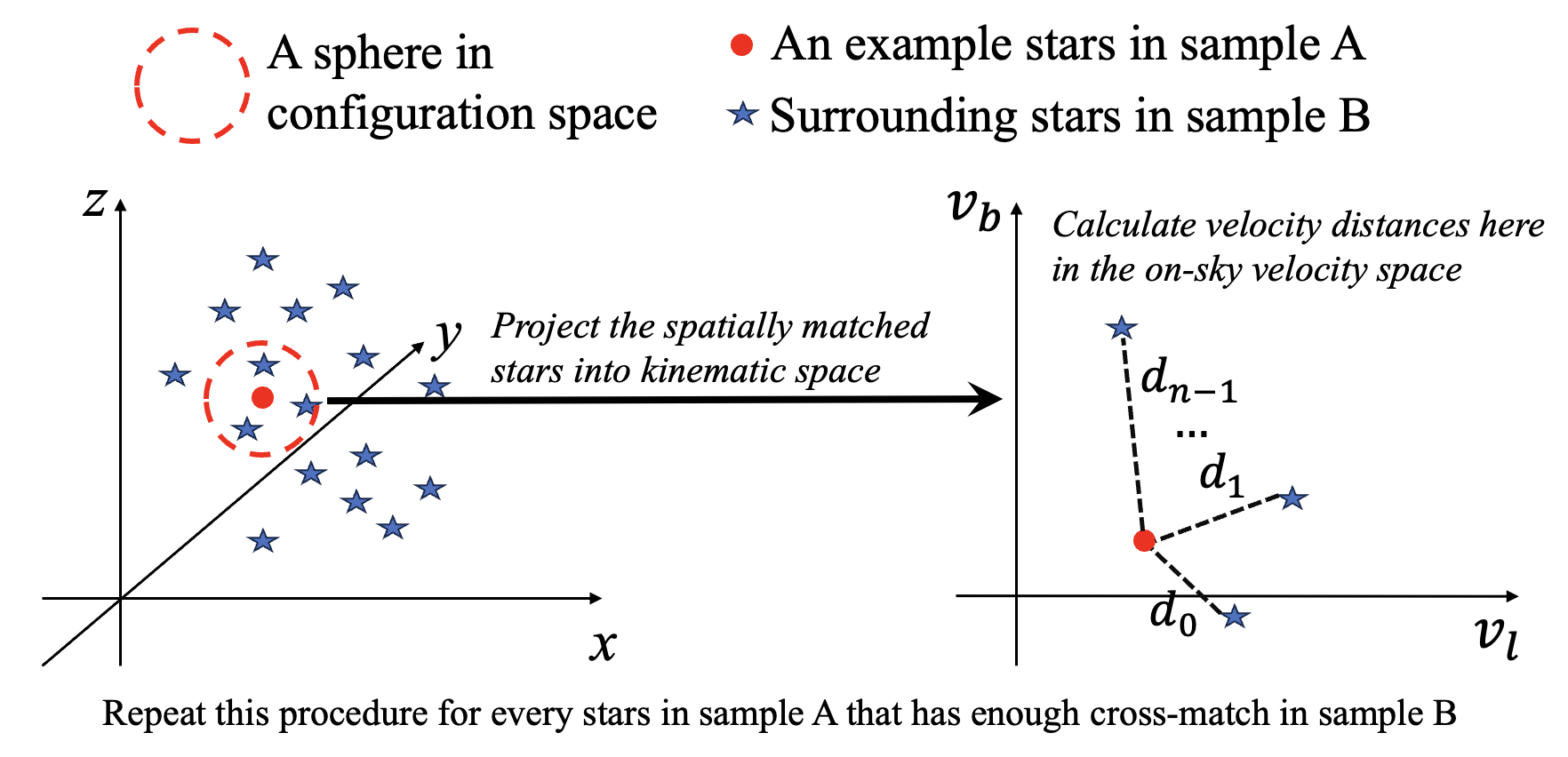}
    \caption{An illustration of the methodology. We first find spatial cross-match from sample B for each stars in sample A using a sphere with radius of $r=0.7$~kpc. Then, the cross-matched stars are projected into the velocity space ($v_\ell$, $v_b$), and the velocity distances are calculated. The procedure is repeated for every star in sample A, and the resulting velocity distance distribution between the sample A and B are recorded as $p_{AB}(\Delta v)$.}
    \label{fig:appendix:cartoon}
\end{figure*}

The procedure of computing the velocity distance distribution between samples A and B, $p_{AB}(\Delta v)$, is illustrated in Fig.~\ref{fig:appendix:cartoon}. The exact same procedure is also used to compute $p_{B'B}(\Delta v)$, and we compare $p_{AB}(\Delta v)$ and $p_{B'B}(\Delta v)$ to derive their kinematic similarity between sample A and B.

\section{Impact from the distance and age uncertainty to the methodology}
\label{Appendix::2PCF_witherr}
\begin{figure*}
    \centering
    \includegraphics[width = \linewidth]{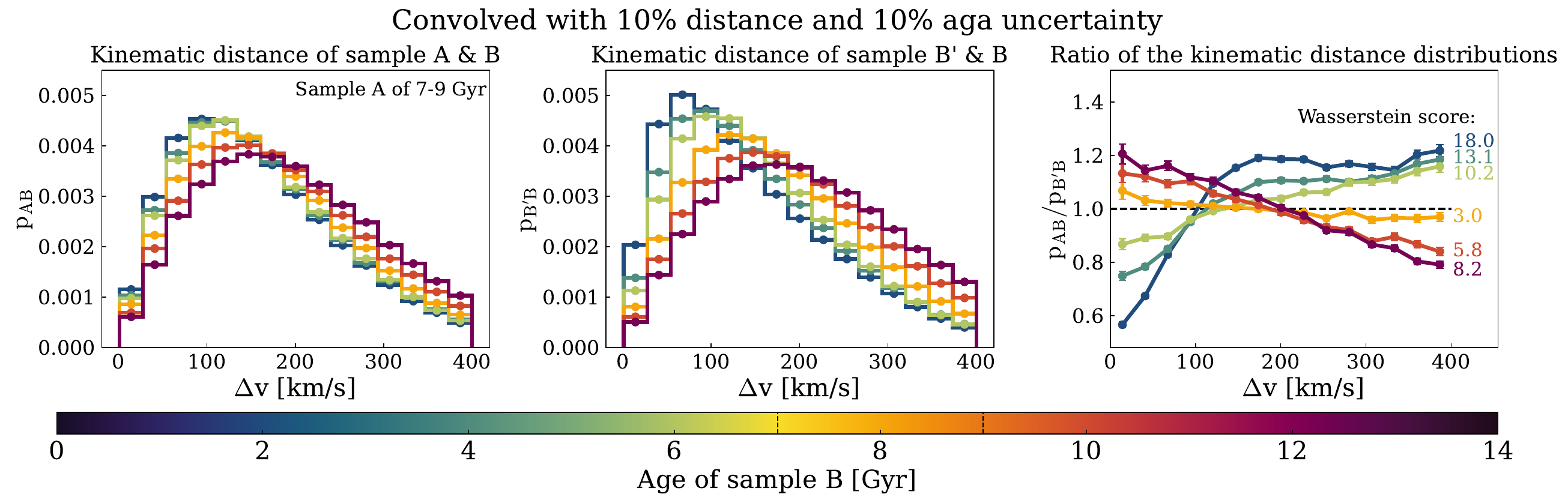}
    \caption{Similar to Fig.~\ref{fig:testing_with_Au18}, but here we re-do it with uncertainty involved. We add $10\%$ distance uncertainty to both samples A and B and $10\%$ age uncertainty to sample B only.}
    \label{fig:testing_with_error_Au18}
\end{figure*}

To verify the validity of the method under realistic observational uncertainty, we repeat our analysis in Section~\ref{subsec:demonstration} but convolve each stellar particle with 10$\%$ distance uncertainty and ensure that the distance uncertainty also propagates to the velocity components ($v_\ell$, $v_b$). To better mimic the Mira variables as in sample B, we also add the $10\%$ age uncertainty to each stellar particle in sample B, comparable to the estimated uncertainties of the age of the Mira variables \citep{ZS23}, which shifts stellar particles in age bins. Similarly to Fig.~\ref{fig:testing_with_Au18}, we show the velocity distance distribution of samples A and B in the left panel, the distribution of samples B' and B in the middle panel, and the ratio between these two distributions in the right panel of Fig.~\ref{fig:testing_with_error_Au18}. In the right panel of Fig.~\ref{fig:testing_with_error_Au18}, we find that the ratio $p_{AB}/p_{B'B}$ is still close to unity for all $\Delta v$ and has the smallest Wasserstein score when the age of sample B matches that of sample A (the orange line), as expected in the error-free case. Compared to Fig.~\ref{fig:testing_with_Au18}, the difference of $p_{AB}/p_{B'B}$ becomes smaller as the age of sample B varies. This is because the age-velocity dispersion relation is blurred by the distances and age uncertainty applied, but noticeable differences in $p_{AB}/p_{B'B}$ persist and also in the Wasserstein score. Hence, we argue that the method is robust under realistic observational uncertainties in RRL and Mira variables. 

We notice that the method generally holds as long as the distance uncertainties in the sample A and B are similar to each other because then the distance uncertainty affects the kinematic space of sample A and B equally. The methodology would become biased if the distance uncertainty between sample A and B differed by more than $5\%$. 

\section{Age, metallicity, mass loss of RR Lyrae from stellar models} \label{app:parsecplot}

\begin{figure}
    \centering
    \includegraphics[trim=0cm 1.6cm 0cm 0cm, clip, width=\linewidth] {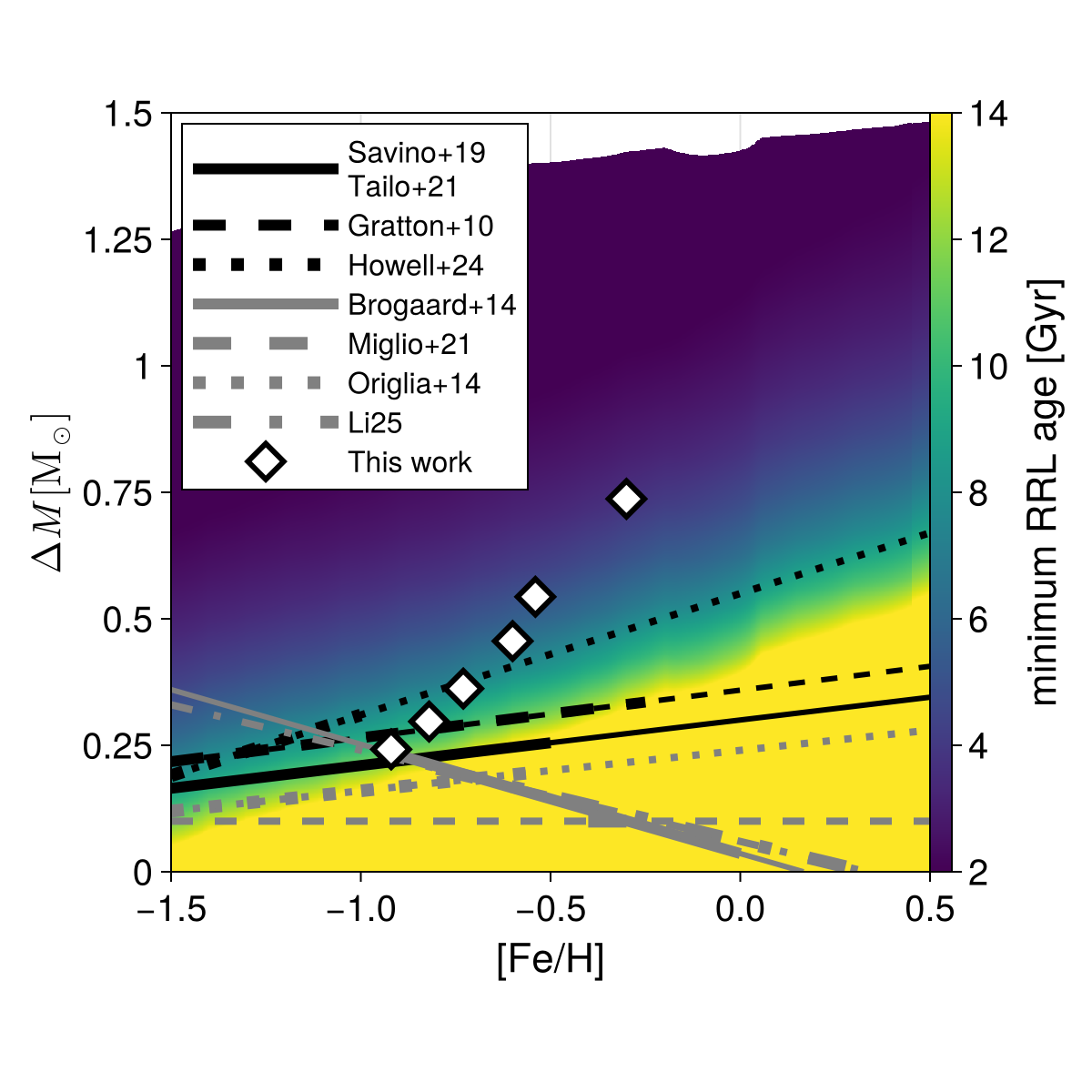}
    \caption{Same as Figure \ref{fig:dmRRL}, bur for the \textsc{basti} stellar evolution models (solar scaled, \citealt{Hidalgo_2018,Pietrinferni_2024}).}
    \label{fig:dmRRL_basti}
\end{figure}

\begin{figure}
    \centering
    \includegraphics[trim=0cm 1.6cm 0cm 0cm, clip, width=\linewidth] {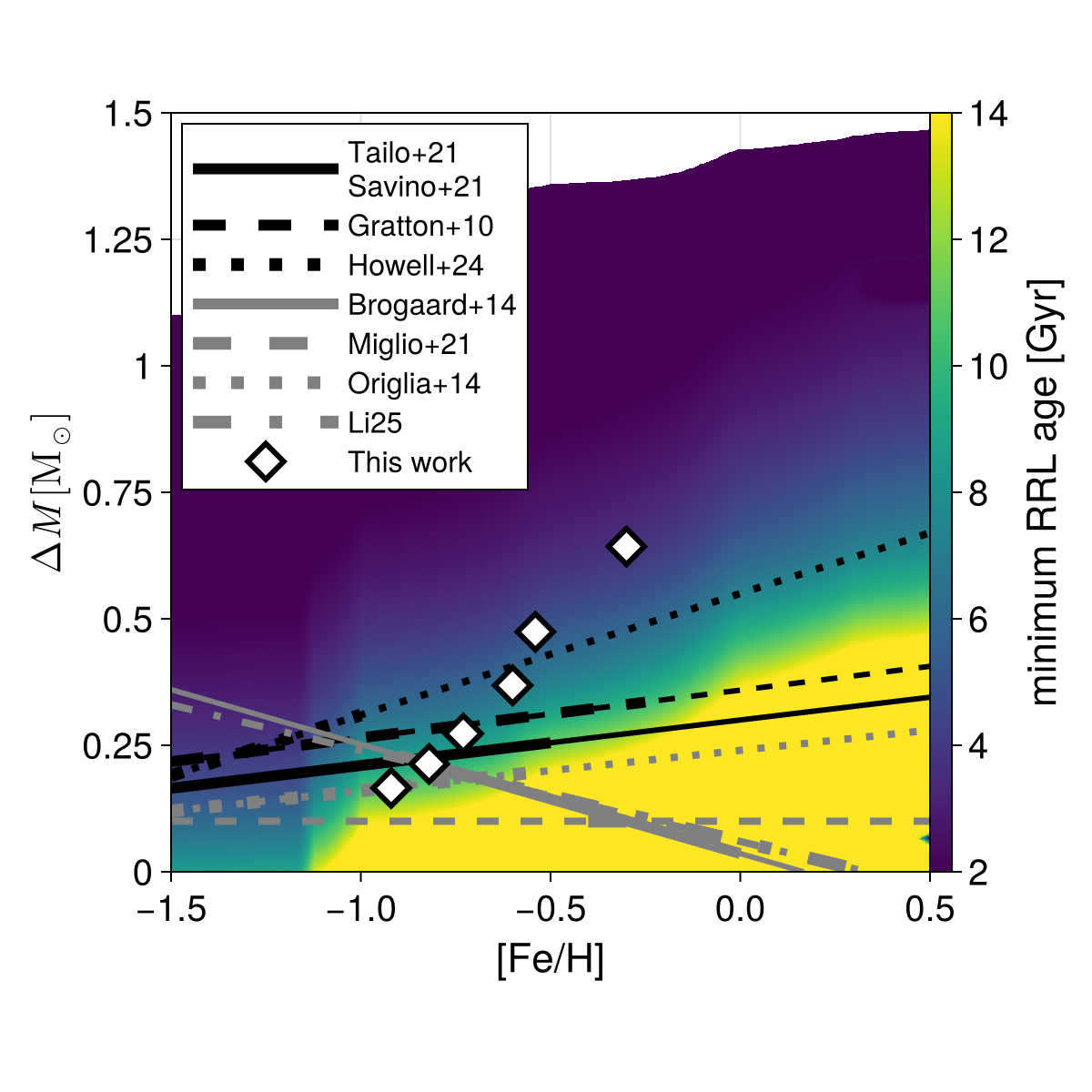}
    \caption{Same as Figure \ref{fig:dmRRL}, bur for the \textsc{Darthmouth} stellar evolution models \citep[$\alpha$-poor models,][]{Dotter_2008}.}
    \label{fig:dmRRL_darth}
\end{figure}

To generate the colour map in Figure \ref{fig:dmRRL} we use  stellar evolution models for low mass stars ($<2 \ \mathrm{M}_\odot$) from the zero age main sequence (ZAMS) up to the Tip of Red Giant Branch (TRGB), and from the zero age horizontal branch (ZAHB)  to the Asymptotic Giant Branch from the \textsc{parsec} V1.2S database\footnote{\url{https://stev.oapd.inaf.it/PARSEC/tracks_database.html
}} \citep{Chen_2015}. 

For each combination of metallicity ($\fehm_\mathrm{i}$) and mass loss ($\Delta M_\mathrm{j}$), we first determine the maximum mass at ZAHB, $M_\mathrm{RRL,max,i}$, predicted by the models at $\fehm=\fehm_\mathrm{i}$.
This mass is selected such that the luminosity and temperature at ZAHB fall within the RRL instability boundaries (IS), as estimated by \cite{Marconi_2015} at the same metallicity.
We further assume as upper limit   $M_\mathrm{RRL,max,i}=0.9 \ \mathrm{M}_\odot$.
This limit has no impact at high metallicity ($\fehm > -1$), but at lower metallicity, massive horizontal branch models (up to 1.5–2 M$_\odot$) become hot enough to enter the instability strip. Whether these stars can actually be observed as RRL variables remains unclear and is beyond the scope of this paper.

Subsequently, we calculate the initial mass of the RRL progenitor as $M_\mathrm{ZAMS,i,j}=M_\mathrm{RRL,max,i} + \Delta M_\mathrm{j}$.
The age, $t_\mathrm{RRL,i,j}$, is then estimated based on the time at the TRGB for models with the initial mass $M_\mathrm{ZAMS,i,j}$ and metallicity $\fehm_\mathrm{i}$.
When the luminosity and temperature values at the ZAHB, and times at the TRGB  are not available in the tabulated models, we employ the \emph{linear} and \emph{rational} interpolation schemes as described in \cite{Iorio_2023} (see their Section 2.1.4). The white regions of Figure \ref{fig:dmRRL} represent models for which $M_\mathrm{ZAMS,i,j} > 2 \ \mathrm{M}_\odot$, and are excluded from this study as these stars do not develop a degenerate helium core and do not undergo a helium flash at the TRGB. 
Given our methodology, the reported RR Lyrae ages  represent a lower limit.

Equivalent results are obtained with the \textsc{Basti}\footnote{\url{http://basti-iac.oa-abruzzo.inaf.it/index.html}} \citep[solar scaled models,][]{Hidalgo_2018,Pietrinferni_2024} 
and \textsc{Darthmouth}\footnote{\url{https://rcweb.dartmouth.edu/stellar/grid.html}} \citep[$\alpha$-poor models,][]{Dotter_2008} stellar evolution models (see Figures \ref{fig:dmRRL_basti}, \ref{fig:dmRRL_darth}).

\section{More results on RRL's age-metallicity correlations}\label{Appendix:Different_ages}

\begin{figure}
    \centering
    \includegraphics[width=0.99\linewidth]{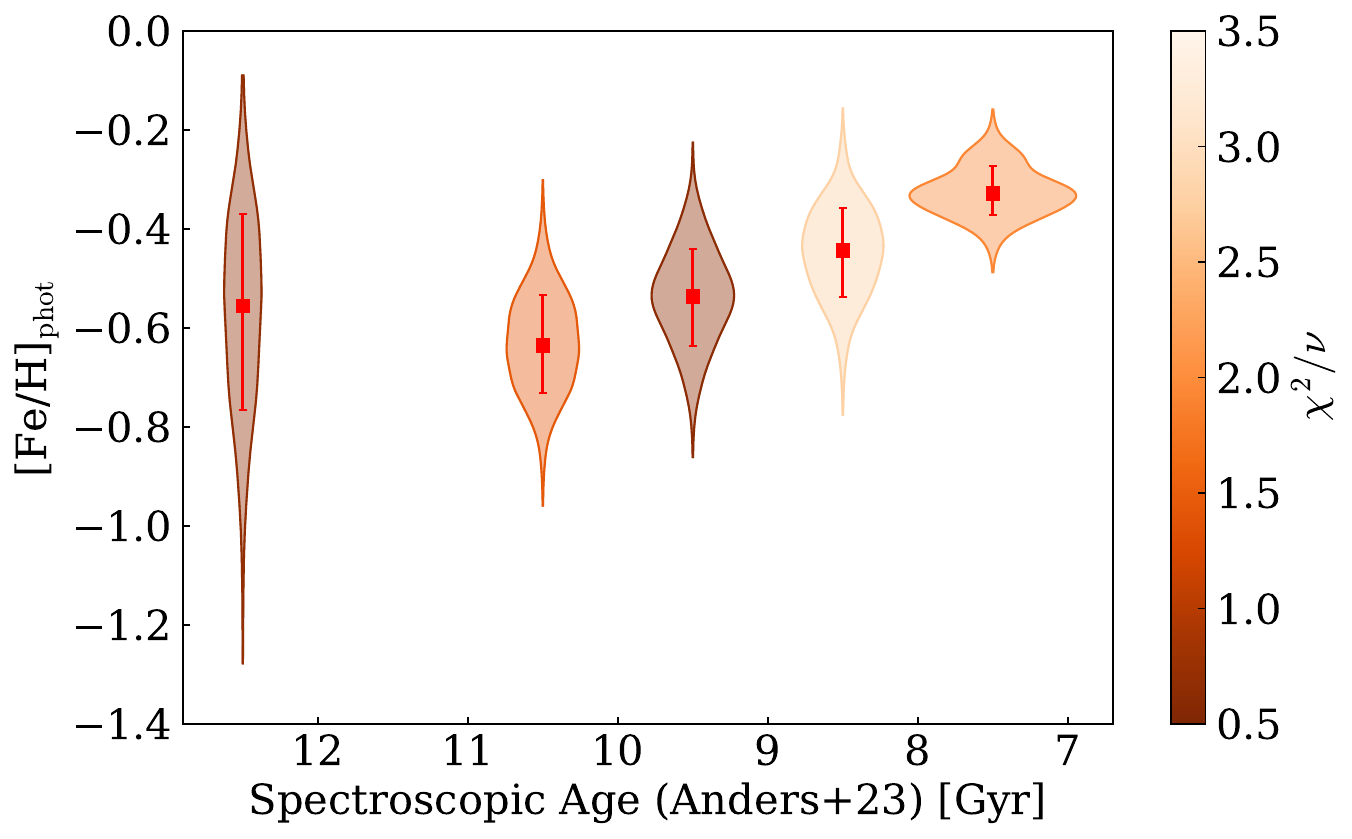}
    \caption{Same as Fig.~\ref{fig:P_feh_given_period}, but we compare RRL kinematics to RGB stars with spectroscopic ages \citep{Anders2023} instead of Mira variables. The results are in broad agreement with the results in Section~\ref{subsec:met_given_age}, but with a slight systematic overestimation of the RRL metallicity at all ages comparing to the results when using Mira variables as the comparison sample.}
    \label{fig:met_given_age_Anders23}
\end{figure}

\begin{figure}
    \centering
    \includegraphics[width=0.99\linewidth]{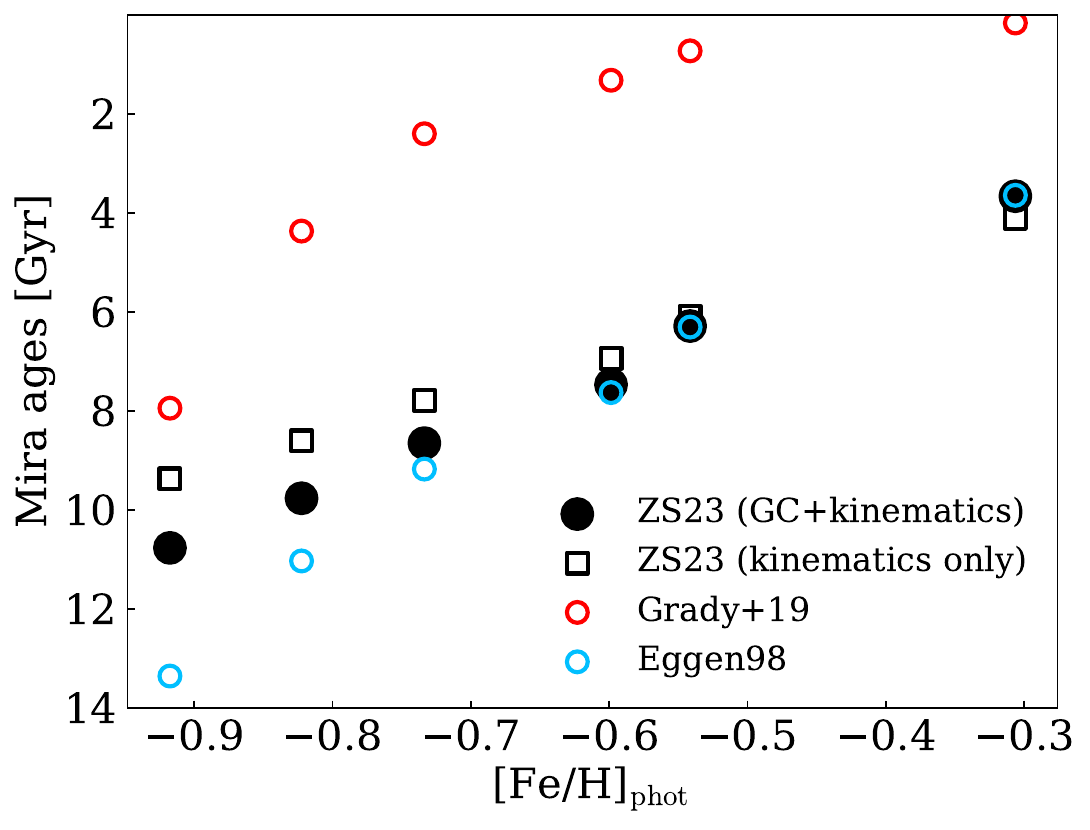}
    \caption{Comparison of the age-metallicity correlations of RRL when different Mira variables period-age relation are adopted. The black dots is the results when we use the period-age relation in \citet{ZS23} calibrated using both Mira kinematics and Mira with GC membership (the relation we mainly used in this work). The black square is when the period-age relation calibrated with pure kinematics inforamtion in \citet{ZS23} is used. The blue circle is when the theoretically-motivated period-age relation in \citet{Eggen1998} is adopted. The red circle is for using the period-age relation calibrated using Mira variables in LMC clusters \citep[][]{Grady2019}.}
    \label{fig:different_Mira_age}
\end{figure}

\begin{figure}
    \centering
    \includegraphics[width=0.99\linewidth]{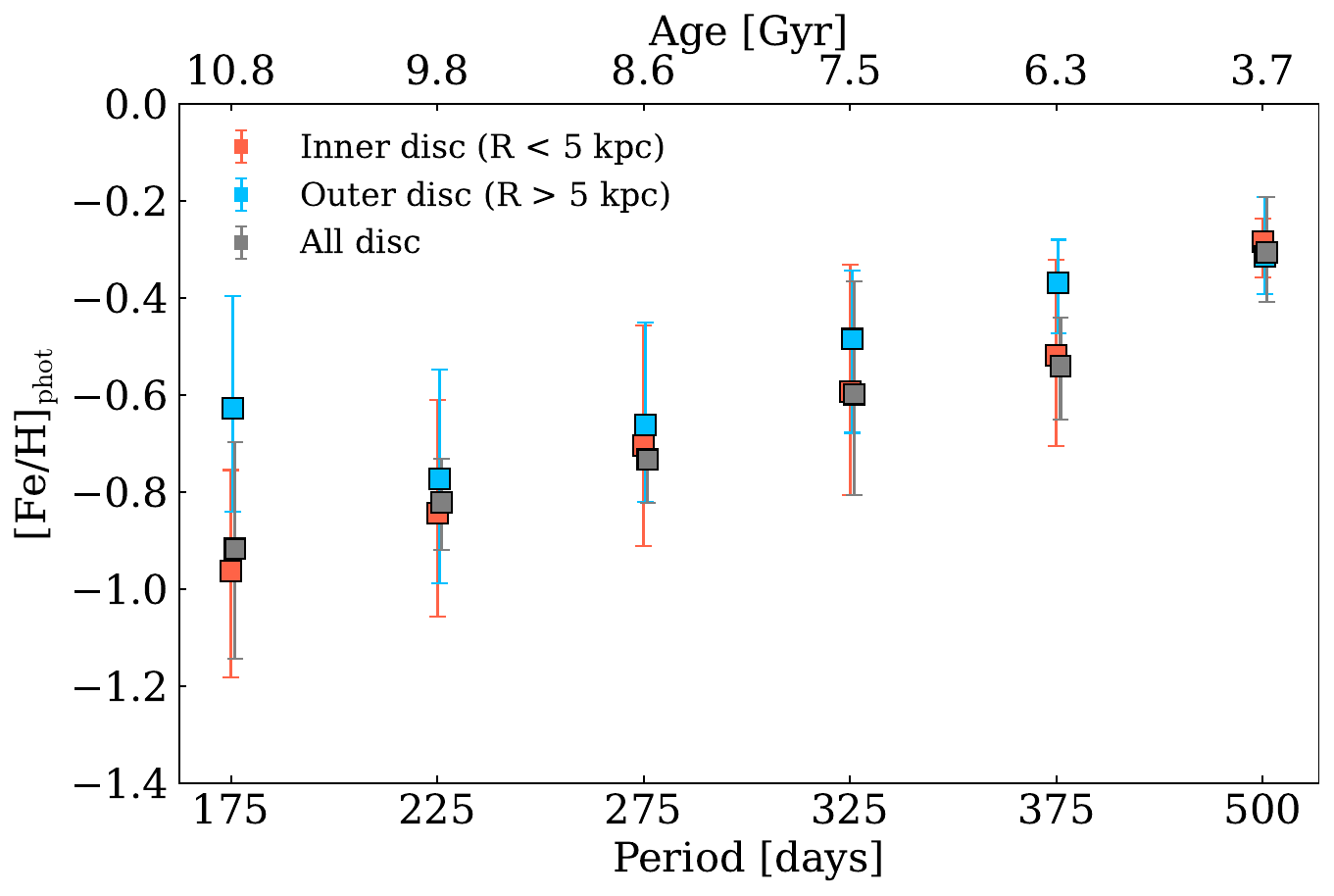}
    \caption{The metallicity-age correlation of the RRLs fitted using Mira variables at the inner (red squares) and outer (blue squares) disc, respectively. The fitting results from the inner and outer discs are systematically different, reflecting the metallicity dependence of the Mira's period-age relation.}
    \label{fig:inner_outer_fit}
\end{figure}

The kinematic comparison algorithm designed in this work requires the compared samples to have a similar spatial coverage. This is the main reason we used Mira variables as the comparison sample, as they have the most similar spatial distribution to other age tracers. For various stellar age estimation techniques, most of them are only valid for stars in the solar vicinity (e.g. the MSTO isochronoal ages and asteroseismic are mainly available for stars within 2 kpc from the Sun), whereas the metal-rich RRL in the adopted sample mainly occupy the inner Milky Way. Among the limited 3836 RRL with $\rm [Fe/H]_{phot}>-1$, only $5\%$ of them are within 2 kpc of the Sun, which is not enough to apply our method. Instead, red giant branch (RGB) stars with ages inferred using spectroscopy cover a larger region in the Milky Way as far as 6-7 kpc from the Sun, where $34\%$ of the metal-rich RRL reside in a similar region. 

To serve as an independent test of our results, we repeat our analysis in Section~\ref{subsec:met_given_age} but using RGB stars with spectroscopic ages (from \citealt{Anders2023}) and distances (from \citealt{Leung2023}) for the comparison sample, as the RGB stars and Mira variables are independent age tracers. The results are shown in Fig.~\ref{fig:met_given_age_Anders23} in a similar way to Fig.~\ref{fig:P_feh_given_period}. The age-metallicity correlation of the metal-rich RRL obtained using the spectroscopic ages is in broad agreement with our results in Section~\ref{subsec:met_given_age} using Mira's ages. The same trend that more metal-rich RRL has younger ages is also seen in Fig.~\ref{fig:met_given_age_Anders23} with a potential bimodal age distribution for RRL with $\rm [Fe/H]>-0.5$. The final metallicity distribution of the RRL obtained by using the spectroscopic age sample is systematically higher than that obtained by using Mira variables at respective ages. It is noteworthy that due to the limited sample size of metal-rich RRL when we use the spectroscopic ages as the comparison sample, the accuracy of our method are expected to drop, and hence, the age-metallicity correlation when using Mira variables as the comparison sample is quantitatively more accurate than that when using RGB stars with spectroscopic ages as the comparison sample. 

As we discussed in Section~\ref{sec:caveats}, the period-age relation of the O-rich Mira variables is still not fully established quantitatively. To illustrate how adopting different period-age relations would affect our results, in Fig.~\ref{fig:different_Mira_age}, we show the resulting age-metallicity correlation of the RRL when we adopt the Mira's period-age relation in \citet{Eggen1998} (blue circle), \citet{Grady2019} (red circle), and \citet{ZS23} (black dot and square). The black dots show the results for using the period-age relation that we adopted throughout this work. The results are similar either when we use the other relation in \citet{ZS23} that used purely kinematic information of the Galactic Mira variables, or when we use the theoretically predicted period-age relation in \citet{Eggen1998}. The period-age relation in \citet{Grady2019} calibrated using the Mira variables in LMC clusters is significantly biased toward a younger age at a fixed Mira's period, which could possibly be attributed to the LMC field star contamination, but other effects cannot be ruled out. Then, the ages of metal-rich RRLs tend to be much younger when we adopt the period-age relation in \citet{Grady2019}, meaning that an even larger mass loss rate is required to produce the observed metal-rich RRLs. However, no matter which period-age relation is adopted, our results confirm the correlation that more metal-rich RRLs are kinematically younger. The ages of RRL with $\rm -1<[Fe/H]<-0.5$ could be as old as $>10.5$ Gyr, considering that Mira variables in globular clusters exhibit a flat age gradient at short-periods \citep{VandenBerg_2013}, but such relation are less likely to be applicable to the field stars because the field Mira (in disc and bulge) demonstrate clear period-kinematic correlation at those periods.

To ensure that the metallicity of Mira variables does not affect our conclusion, we split the RRL and Mira variable sample into the inner ($R<5$~kpc) and outer ($R>5$~kpc) disc region, respectively. We repeat the same analysis on the inner and outer disc samples, and the results are shown in Fig.~\ref{fig:inner_outer_fit}. We find a systematic difference in the resulting age-metallicity correlation of RRL in the inner disc and in the outer disc regions, where the fitted metallicity in the outer disc sample is higher than that in the inner disc. This could be attributed to the metallicity dependence of the Mira's period-age relation \citep{FrogelWhitelock1998, FeastWhitelock2000}. Due to the radial metallicity gradient along the disc \citep{Braganca2019}, the ages of the inner and outer disc Mira variables are different at each period. However, as the metallicity dependence of Mira variables' period-age relation has not yet been quantified, and the difference in the fitting results is rather minor, we treat this as a systematic uncertainty of the work and leave this for future work.

\section{Extra figures}

\begin{figure*}
    \centering
    \includegraphics[width=0.99\linewidth]{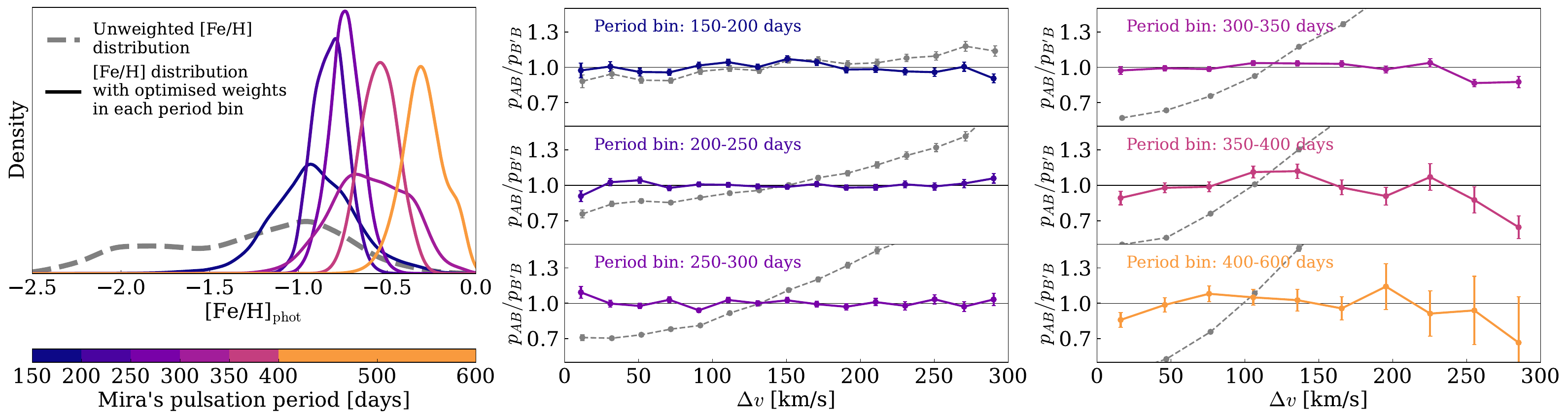}
    \caption{Similar to Figure.~\ref{fig:results_age_distribution} but to demonstrate the kinematic similarity between the weighted kinematic distribution of RRL and the Mira variables at the corresponding age bins as shown in Figure.~\ref{fig:P_feh_given_period}. {\it Left:} the optimised metallicity distribution of RRL for Mira variables with different ages. The grey dashed line shows the distribution of the unweighted RRL metallicity distribution in the sample. {\it Right panels:} the coloured line shows the ratio of the kinematic distribution between the Mira variables and the weighted RRLs, which they are all consistent with unity. The grey dashed line shows the ratio when comparing Mira variables to the unweighted RRL distribution.}
    \label{fig:met_given_age_kinematic_histogram}
\end{figure*}









\bsp	
\label{lastpage}
\end{document}